\documentclass[10pt,review]{elsart}
\usepackage[utf8]{inputenc}
\usepackage{graphicx}
\usepackage{amsmath,scalefnt}
\usepackage{amssymb}
\usepackage{amsfonts}
\usepackage{xcolor}
\usepackage[numbers]{natbib}
\usepackage{setspace}
\newcommand{\ord}{{O}}
\newcommand{\IN}{{\textrm{\tiny IN}}}
\newcommand{\GC}{{\textrm{\tiny GC}}}
\newcommand{\optS}{{\textrm{S}}}
\newcommand{\optR}{{\textrm{R}}}
\newcommand{\Id}{{\boldsymbol{I}}}
\newcommand{\CapText}{\it \small }
\usepackage{multirow}
\newfont{\notapolice}{cmss10}
\begin{document}
\begin{frontmatter}
\title{Very high-order Cartesian-grid finite difference method on arbitrary geometries}
\author[fisiks,math]{S.~Clain,}
\ead{clain@math.uminho.pt}
\author[fisiks]{D.~Lopes,}
\author[fisiks,math]{R.~M.~S.~Pereira,}
\address[fisiks]{Centre of Physics, Campus de Gualtar, 4710 - 057 Braga, Portugal}
\address[math]{Department of Mathematics,\\ University of Minho, Campus de Azur\'em,\\ 4080-058 Guimar\~aes, Portugal}


\begin{abstract}
An arbitrary order finite difference method for curved boundary domains with Cartesian grid is proposed. The technique handles in a universal manner Dirichlet, Neumann or Robin condition. We introduce the Reconstruction Off-site Data (ROD) method, that transfers in polynomial functions the information located on the physical boundary. Three major advantages are: (1) a simple description of the physical boundary with Robin condition using a collection of points; (2) no analytical expression (implicit or explicit) is required, particularly the ghost cell centroids' projection are not needed; (3) we split up into two independent machineries the boundary treatment and the resolution of the interior problem, coupled by the the ghost cell values. Numerical evidences based on the simple 2D convection-diffusion operators are presented to prove the ability of the method to reach at least the 6th-order with arbitrary smooth domains.
\end{abstract}
\end{frontmatter}

\section{Introduction}
Most real problems take place on arbitrary geometries and one has to account for the boundary complexity to reproduce, at the numerical level, the interactions between the interior problem and the boundary conditions. Boundary layer and turbulence are among others, examples of phenomena that are mostly driven by the boundary condition and attention would be drawn on the numerical schemes to provide the correct behaviour of the numerical approximation. Very high order methods (we mean strictly higher than the second-order of approximation) turned out to be an excellent tool in capturing the local geometry details and improving its accuracy. The counterpart is that additional efforts have to be made to treat a domain with curved boundaries. Indeed, popular schemes are usually restricted to, at most, the second-order case when boundary conditions are not exactly localised on the nodes of the grid and the edge of cells. 

Several techniques have been recently developed in the unstructured mesh context to preserve the optimal order. We refer to \cite{CLN2019,RNC2019,FFCR2020} for a recent review. In the present study, we are focusing on the specific case of the finite difference method on Cartesian grids. It is a very popular discretisation technique due to the low data storage, free underlying structures, and draws some advantages due to the simplicity of the numerical schemes \cite{LZH2009}. Since the beginning of the seventies, and after the pioneer paper of Peskin \cite{Pe1972}, finite difference method with the boundary embedded in a Cartesian grid provides superior advantages over the conventional boundary-conformal approach since the computational mesh remains unchanged with respect to the boundary. 

Historically, Immersed Boundary (IB) methods were classified into two categories: continuous force and discrete force approach (see \cite{MiIa2005,IaVe2003} for a detailed overview). Nowadays, such a classification turns to be obsolete and the discrete force approach falls into a general framework that consists in transferring information located on the boundary into information supported by some nodes of the grid. Introduced in the original work of Mohd-Yosuf \cite{MY1997} and extended by the so-called ghost cell method \cite{FAM1999}, several authors have contributed to improve the accuracy and stability of the technique \cite{VMOH2000,FVOM2000,MID2001}. Roughly speaking, a set of cells tagged ghost cells are identified around the computational domain. For each ghost cell of centroid $M$, the orthogonal projection point on the physical boundary $P$ is determined together with the normal vector $n$. We define the image point $BI$ in the physical domain by symmetry and a value is assigned using linear, bi-linear or quadratic reconstructions involving neighbouring points \cite{TF2003,CCKR2018}. Then a simple extrapolation of the BI and $P$ values transfers the Dirichlet or Neumann condition into a equivalent Dirichlet condition at the ghost cell centroid $M$. Extension using several points on the semi-line $(M,n)$ have been proposed to provide a second-order approximation \cite{GSB2003,NL2014,KBF2017} with the Neumann condition and fourth/fifth order reconstruction along the normal have been recently proposed \cite{ApPe2012,BMZ2016}. 

All the previous methods deal with a two-steps strategy. First, several approximations are computed with nodes interpolation at interior points located  along the normal line. Second, an extrapolation (ghost point) or an interpolation (boundary point) using the boundary condition is computed at the ghost cell centroid. An alternative approach consists in performing both the nodes and boundary interpolation in one step by computing the best polynomial function in the least squares sense. The introduction of boundary condition with a polynomial representation dates back to the papers of \textit{Tiwar and Kuhnert (2001)} \cite{TK2001,TK2002} in the pointset method context while the same idea was independently proposed by \textit{Ollivier-Gooch and Van Altena (2002)} \cite{GoVa2002} for the finite volume method. A similar technique was proposed for Cartesian grids in \cite{LMZ2008,SeMi2011,XLF2014}. In ever, the boundary condition is taken into account in a weak sense, \textit{i.e.} with a weighted least squares method, hence the resulting polynomial function does not exactly satisfy the condition, but up to the reconstruction order.

We propose a different strategy to include the boundary condition located on the physical domain while preserving the optimal (very) high-order. The overall picture of the problem is, on the one hand, that the boundary data is not situated on the nodes and one has to transfer the information from the physical domain onto the computational domain. Secondly, the boundary condition treatment presents a high level of independence regarded to the interior problem. The main idea consists in elaborating a mathematical object (a polynomial function for instance) that catches all the information about the location and the boundary condition we shall insert in the numerical scheme. For example, in the finite volume context, the boundary information is converted into flux on the interface edges  \cite{CLN2019,RNC2019}. We tag the method "Reconstruction Off-site Data" (ROD) to highlight the transfer of information located on the physical boundary and not on the grid (Off-site Data) into a polynomial (Reconstruction). We generalise the concept with a tidy separation between the boundary treatment and the numerical scheme and adapt the technique to the Cartesian grid context using the ghost cell method. 

Noticeable differences between our method and the other authors will be mentioned. A strict inclusion of the boundary condition is obtained by imposing the polynomial reconstruction to exactly satisfy the boundary condition. Moreover, the routine treats the Dirichlet and Neumann conditions as a particular case of the Robin condition without any specificity. This provides a universal framework to deal with all kind of conditions. We also stress that our method does not require the ghost cell centroid projection onto the physical boundary and it just needs a list of points that belong to the frontier. At last, the method is of arbitrary order for arbitrary geometries, depending on the polynomial degree involved in the reconstruction on the local regularity of the border. As a final note, we highlight that we restrict the study to the simple convection diffusion problem on purpose for the sake of simplicity in order to focus on the main objective of the present work: the treatment of boundary condition on arbitrary geometries.

The organisation of the paper is the following. We present in section 2 the equations and the numerical methods that we consider in this paper. 
The details of the Reconstruction Off-site Data procedure is explained in section 3 while section 4 is dedicated to a new ADI strategy to solve the interior problem. Section 5 is dedicated to the coupling between the interior and boundary problem and propose a study on an accelerated fix-point solver. Some numerical results that are obtained with this methodology are presented in section 6 to check the accuracy and computational effort.

\section{Convection diffusion on curved boundary domain}
We introduce the basic ingredients to deal with the discretisation of the equations and the boundary. In particular, we define the solver operator that deals with the discrete convection diffusion equation, deriving form the standard finite difference method, and the boundary operator, involving a very simple discrete approximation of the boundary. 
\subsection{Domain discretisation}
Let $\Omega\subset \mathbb R^2$ be an open bounded set. We consider the linear scalar convection diffusion problem for the present study: find a function $\phi$ on $\Omega$ such that
\begin{equation}\label{eq:ConvDiff}
\mathfrak F(\phi)=f-\nabla \cdot (U\phi-\kappa\nabla \phi)=0, \quad \textrm{in } \Omega,    
\end{equation}
with $\kappa\geq 0$, $U=(u_x,u_y)$, equipped with the boundary condition 
\begin{equation}
\label{eq:BoundCond}
\mathfrak B(\phi)=g-\alpha \phi-\beta \nabla\phi\cdot n=0, \quad \textrm{on } \partial\Omega,    
\end{equation}
with $g$ a given function on the boundary $\partial \Omega$, $\alpha$, $\beta$ are real numbers and $n$ is the outward normal vector on $\partial \Omega$.

\begin{figure}[ht]
\centering
\includegraphics[width=0.5\textwidth]{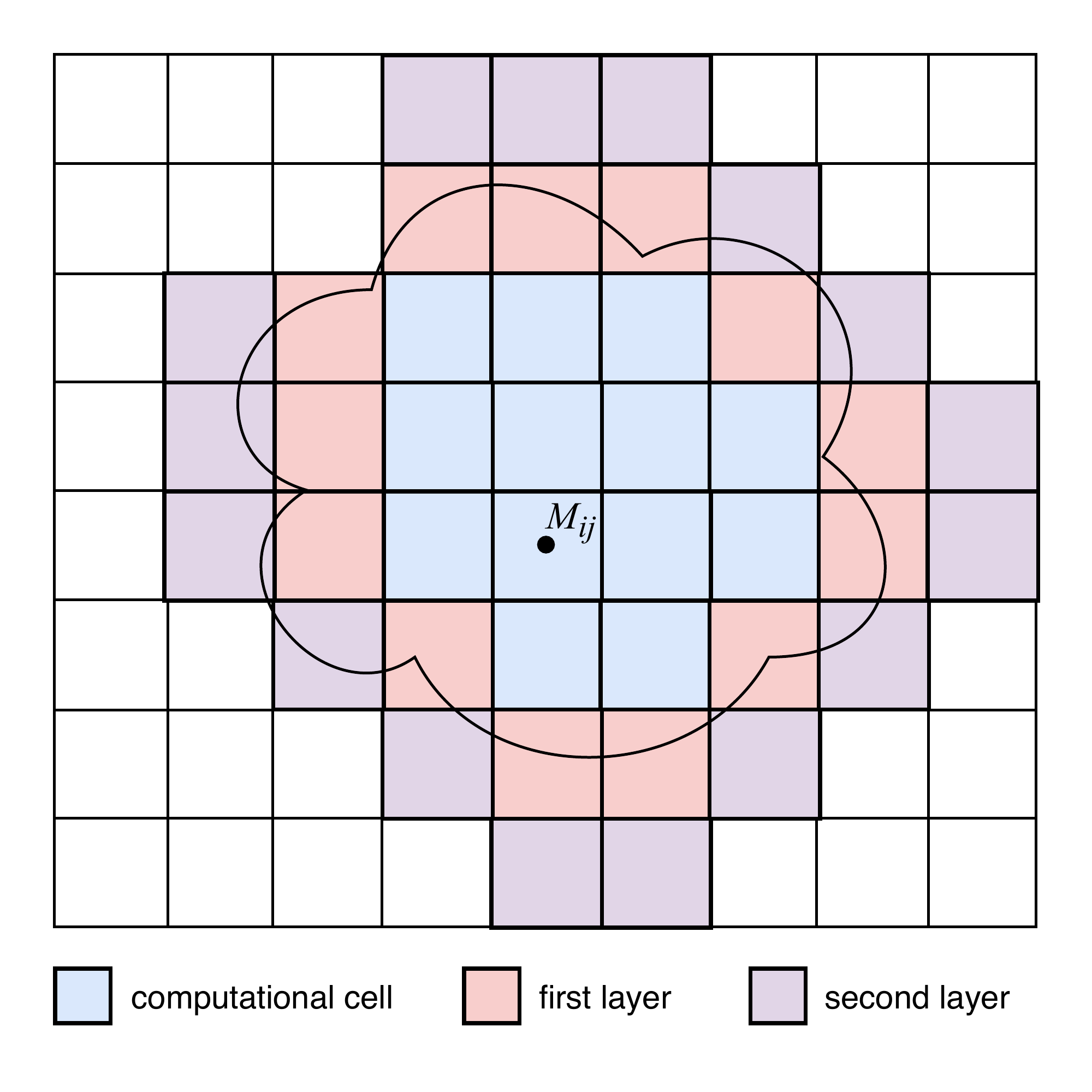}
\caption{Example of a numerical domain and two layers of ghost cells}
\label{fig:grid}
\end{figure}

We denote by $\Lambda=[0,L_x]\times [0,L_y]\subset \mathbb R^2$ the rectangle that contains the sub-domain $\Omega\subset \Lambda$ with Lipschitz, smooth piece-wise, boundary $\partial \Omega$. For $I$, $J$, two given integer numbers, we set $\Delta x=L_x/I$, $\Delta y=L_y/J$ the mesh sizes.  We adopt the following notations with $i=0,\cdots,I-1$ and $j=0,\cdots,J-1$:
\begin{eqnarray*}
& &x_{i-1/2}=i\Delta x,\quad x_i=x_{i-1/2}+\Delta x/2,\quad x_{i+1/2}=x_{i-1/2}+\Delta x, \\
& &y_{j-1/2}=j\Delta y,\quad y_j=y_{j-1/2}+\Delta y/2,\quad y_{j+1/2}=y_{j-1/2}+\Delta y, \\
& & C_{i,j}=[x_{i-1/2},x_{i+1/2}]\times [y_{j-1/2},y_{j+1/2}].
\end{eqnarray*}
Moreover, $N^1_{i,j}$, $\cdots$, $N^4_{i,j}$ stand for the four nodes of cell $C_{i,j}$ while $M_{i,j}=(x_i,y_j)$ is the centroid. Dropping the indices $i$, $j$, the simpler notation $N^{1},\cdots,N^{4}$ and $M$ is used when the cell is clearly identified. The mesh $\mathcal M_\Delta=\mathcal M_\Delta(\Lambda)$ gathers the cells of domain $\Lambda$ and we define the grid associated to domain $\Omega$ by
$$
\mathcal{M}_{\Delta}(\Omega)=\big \{C \in \mathcal{M}_\Delta, \textrm{ such that }\ N^1,\cdots, N^4\in \Omega\big \}, \qquad
\Omega_\Delta=\bigcup_{C\in \mathcal{M}_{\Delta}(\Omega)} C,
$$
where $\Omega_\Delta$ stands for the numerical domain. Cells $C\in\mathcal{M}_{\Delta}(\Omega)$ are tagged active cells since they correspond to the computational domain (see figure \ref{fig:grid}).

To define the ghost cells, we introduce the rook distance between two cells with
$$
d_r(C_{i,j},C_{i',j'})=\left \{
\begin{array}{l}
|i-i'| \textrm{ if } j=j',\\
|j-j'| \textrm{ if } i=i',\\
+\infty \textrm{ otherwise},
\end{array}\right .
$$
and the distance between a cell and domain $\Omega_\Delta$ with
$$
d_r(C_{i,j},\Omega_\Delta)=\min_{C_{i',j'}\in\mathcal M_\Delta(\Omega)} d_r(C_{i,j},C_{i',j'}).
$$
The first layer of ghost cells is then characterised by cells with a distance to $\Omega_\Delta$ is equal to $1$ , namely
$$
\mathcal L_1(\Omega_\Delta)=\{ C_{i,j}\in\mathcal M_\Delta,\ d_r(C_{i,j},\Omega_\Delta)=1\}.
$$
Straightforward  extensions is made for the second layer $\mathcal L_2(\Omega_\Delta)$ and third layer $\mathcal L_3(\Omega_\Delta)$ of ghost cells.

Finally, for a given cell $C_{i,j}$ and $\ell\in\mathbb N$, we define the $\ell$-stencil $\mathcal V_{\ell}(C_{i,j})\subset \mathcal M_\Delta(\Omega)$, as the list of the $\ell$-closest cells of $\mathcal M_\Delta(\Omega)$ to $C_{i,j}$. Notice that a stencil is only constituted of cells from the computational domain.

\subsection{Boundary discretisation}
\begin{figure}[ht]
\centering
\includegraphics[width=0.5\textwidth]{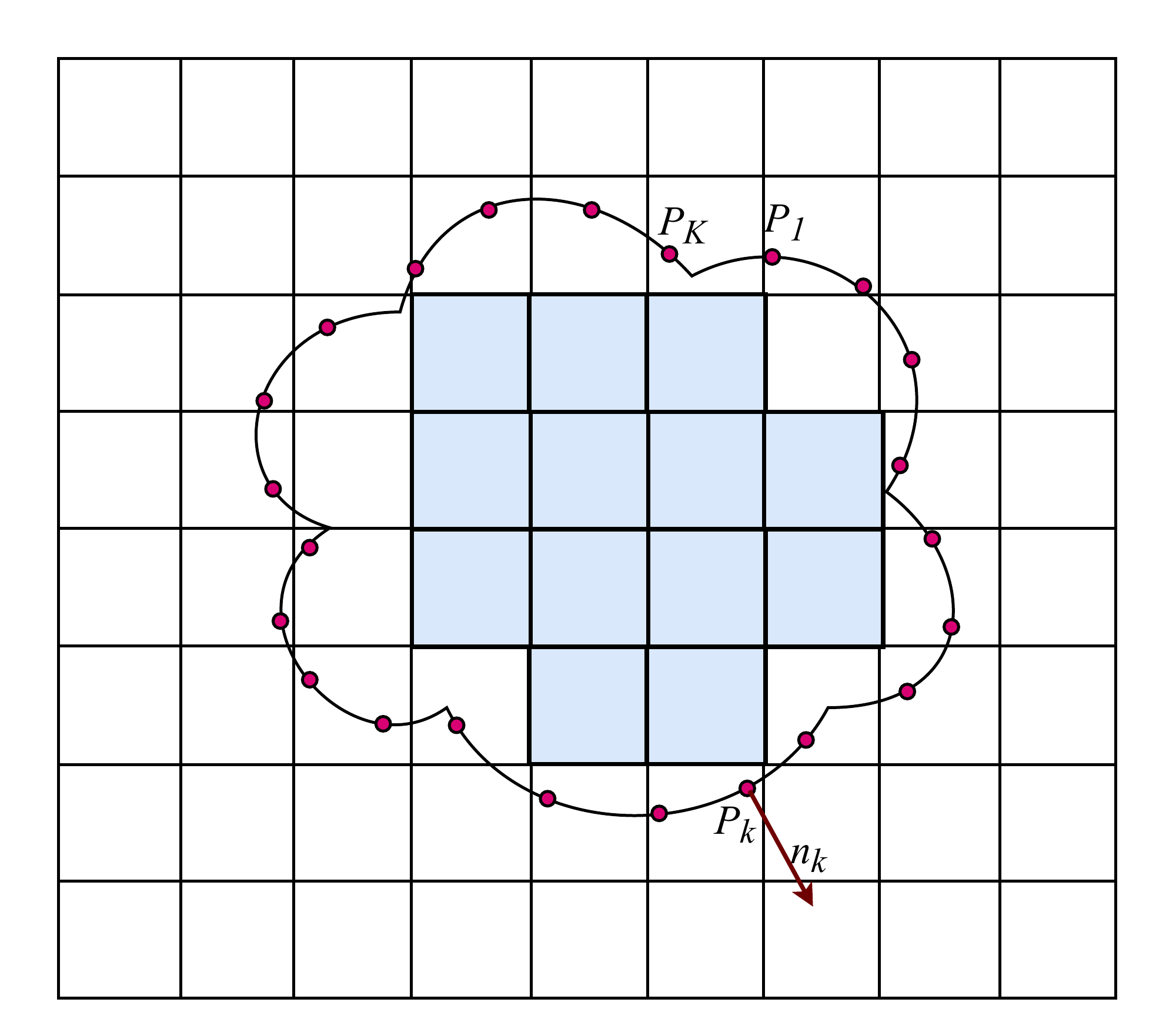}
\caption{Example of a collar on the physical boundary}
\label{fig:collar}
\end{figure}

To handle the boundary at the discrete level, we consider a set of $K$ points $P_k\in\partial \Omega$, $k=1,\cdots,K$ we denote the collar as $\mathcal C_\Delta$ (see figure \ref{fig:collar}). We only assume that $|P_{k}P_{k+1}|=\ord(\Delta s)$, where $P_{K+1}=P_1$ by convention and $\Delta s$ is the characteristic length of the collar. We shall take $\Delta s=\sqrt{\Delta x\Delta y}$ to guarantee the same accuracy than the numerical schemes. In addition to listing the positions, the collar also provides the outward normal vector $n_k$ of the physical boundary at points $P_k$. 
\begin{rem}
There are several methods to compute the collar point in function of the boundary characterisation depending on the description of the boundary: Jordan parametric curve, level set function, polar coordinate curve. We refer to \cite{FFCR2020} for a detail presentation. Nevertheless, we do not require any analytical description of the boundary to perform the method except a list of points and normal vectors. In particular, orthogonal projection of the ghost cell centroid is not required.
\end{rem}

\subsection{Reconstruction and solver operators}
Before presenting the technical aspects, we give a general view of the method by defining the main operators. 
Let $\phi=\phi(x,y)$ be a function defining on the domain $\Omega$. We denote by $\boldsymbol{\Phi}[i,j]$ the approximations of $\phi(M_{i,j})=\phi(x_i,y_j)$ for $C_{i,j}\in \mathcal M_\Delta(\Omega)$ and the ghost cell values for $C_{i,j}\in \mathcal L_m(\Omega_\Delta)$, $m=1,2,3$ that we gather in matrix $\boldsymbol{\Phi}\in \mathbb R^{I\times J}$, the other entries still undetermined. Adopting a new one-index  numbering $(i,j)\to \ell$, matrix $\boldsymbol{\Phi}$ gives rise to two vectors: $\boldsymbol{\Phi}_{\IN}$ corresponds to the values on the active cells \textit{i.e.} the cells that belong to $\mathcal{M}_{\Delta}(\Omega)$, while $\boldsymbol{\Phi}_{\GC}$ gathers the values on ghost cells of the different layers. The method we propose is based on two operators coupling the active cells and the ghost cells. 

The linear reconstruction operator (ROD operator)
\begin{equation}
\label{eq:Roperator}
\boldsymbol{\Phi}_{\IN}\to \boldsymbol{\Phi}_{\GC}=\optR(\boldsymbol{\Phi}_{\IN};g)
\end{equation}
that provides the ghost cell values, given the active cell values and the boundary condition. 
We define the linear solver operator 
\begin{equation}
\label{eq:Soperator}
\boldsymbol{\Phi}_{\GC}\to \boldsymbol{\Phi}_{\IN}=\optS(\boldsymbol{\Phi}_{\GC};f)
\end{equation}
that provide the approximation on the active cells, given the values on the ghost cells and the right-hand side function.
The numerical solution of the convection diffusion problem (\ref{eq:ConvDiff})-(\ref{eq:BoundCond}) satisfies both the conditions
$$
\boldsymbol{\Phi}_{\GC}=\optR(\optS(\boldsymbol{\Phi}_{\GC};f);g),\qquad \boldsymbol{\Phi}_{\IN}=\optS(\boldsymbol{\Phi}_{\GC};f),
$$
and suggest an iterative method to reach the fix-point solution. Another approach consists in introducing the global residual operator
$$
\mathcal S(\boldsymbol{\Phi}_{\IN},\boldsymbol{\Phi}_{\GC})=\left (
\begin{array}{c}
\optS(\boldsymbol{\Phi}_{\GC};f)-\boldsymbol{\Phi}_{\IN}\\
\optR(\boldsymbol{\Phi}_{\IN};g)-\boldsymbol{\Phi}_{\GC}
\end{array}
\right )
$$
such that the numerical solution is given by $\mathcal S(\boldsymbol{\Phi}_{\IN},\boldsymbol{\Phi}_{\GC})=0$. The second approach provides a matrix-free linear operator that could be handle by an iterative method of type GMRES or BiGCStab.

\section{The Reconstruction of Off-site Data method (ROD)}
We detail the polynomial reconstruction operator (\ref{eq:Roperator}) to provide a high accurate approximations of the solution in the ghost cell for smooth curved boundaries. We recall that the frontier is only characterised by a simple list of points and normal vectors. The major difference with the reconstruction proposed in \cite{GoVa2002} is the specific treatment of the boundary condition. Indeed, we combine a least squares method over a stencil of active cells but imposing the polynomial to satisfy the general Robin condition on two collar points. In other words, we apply the least squares procedure to a convex subset of polynomials that strictly respect the boundary condition.  

\subsection{The constraint optimisation problem}
We assume that the entries of $\boldsymbol{\Phi}_{\IN}$ contain an approximation of $\phi$. Let $C_{i,j}$ be a ghost cell. We adopt the multi-index notation $\gamma=(\gamma_x,\gamma_y)$ and for any points $X=(x,y) \in\mathbb R^2$, we define the polynomials $\pi\in\mathbb P_d(\mathbb R^2)$ centred at the centroid $M_{i,j}$ as
$$
\pi(X)=\pi(X;M_{i,j})=\sum_{|\gamma|\leq d} a_\gamma \left (\frac{x-x_i}{\Delta x}\right )^{\gamma_x}\left (\frac{y-y_j}{\Delta y}\right )^{\gamma_y}.
$$
The lexicographic order given by $(0,0)$, $(1,0)$, $(0,1)$, $(1,1)$, $(2,0)$, $\cdots$ defines a one-to-one function $m\to \gamma=\boldsymbol{\gamma}[m]$ where $m$ is the rank in the lexicographic order (for instance $\boldsymbol{\gamma}[8]=(3,2)$). Polynomial of degree $d$ is constituted of $c_d=d(d+1)/2$ monomial functions we rewrite under the compact form $\pi(X;\boldsymbol{a})=\boldsymbol{a}\cdot \boldsymbol{\chi}_{i,j}(X)$ where vector $\boldsymbol{a}\in\mathbb R^{c_d}$ collects the polynomial coefficients $\boldsymbol{a}[m]=a_\gamma$, $\gamma=\boldsymbol{\gamma}[m]$ while $\boldsymbol{\chi}_{i,j}(X)$ collects the monomial functions.
Finally, we define the energy functional
$$
\mathfrak E_{i,j}(\pi;\mathcal V,\boldsymbol{\Phi})=\sum_{M_{i',j'}\in\mathcal V} \frac{1}{2}\Big (\pi(M_{i',j'})-\boldsymbol{\Phi}[i',j']\Big)^2
$$
that represents the quadratic error between the polynomial representation centred at $M_{i,j}$ and the approximations over the stencil $\mathcal V=\mathcal V(C_{i,j})$ constituted of active cells.

\begin{figure}[ht]
\centering
\includegraphics[width=0.45\textwidth]{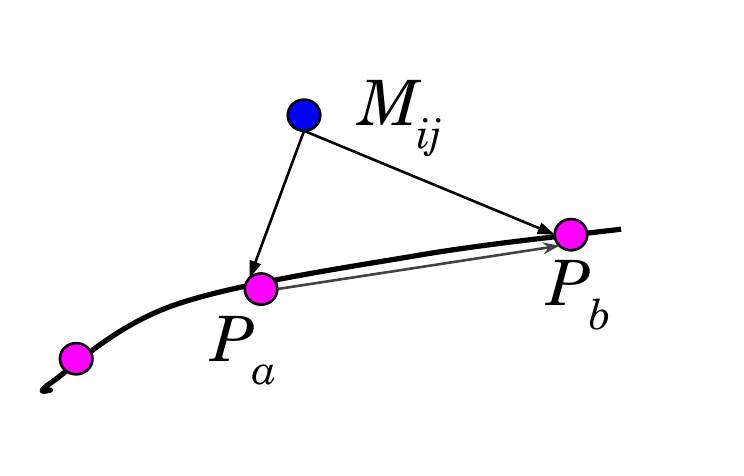}
\caption{The cone condition select the two point of the collar where the boundary condition will be prescribed in the ROD reconstruction}
\label{fig:cone_condition}
\end{figure}

We denote by $P_a$ the closest point of the collar, {\it i.e.}
$$
P_a=P_{a(i,j)}=\arg \min_{P_k \in \mathcal C(\Omega)}|PM_{i,j}|
$$
where we drop the indices $i,j$ for the sake of simplicity. Then there exist two collar points $P_L$, $P_R$ on both sides of $P_a$ and we choose one of the two points, denoted $P_b=P_{b(i,j)}$ that satisfies the cone condition (see Figure \ref{fig:cone_condition})
$$
P_a M_{i,j}\cdot P_a P_b >0, \qquad P_b M_{i,j}\cdot P_a P_b>0.
$$

The Reconstruction of Off-site Data method consists in seeking the coefficients $\boldsymbol{a}_{i,j}$ of the polynomial $\widehat \phi_{i,j}(X)=\boldsymbol{a}_{i,j}\cdot \boldsymbol{\chi}_{i,j}(X)$ such that
$$
\widehat \phi_{i,j}=\arg \min_{\pi \in\mathbb P^d} \mathfrak E_{i,j}(\pi;\mathcal V,\boldsymbol{\Phi})
$$
under the restriction 
$$
\mathfrak B_{i,j}^{r}(\pi)=\alpha(P_r)\pi(P_r)+\beta(P_r)\nabla_X \pi(P_r)\cdot n_{r}-g(P_r)=0,\quad  r=a,b.
$$

\subsection{Calculation of the ROD polynomial}
To determine the solution, we define the Lagrangian functional
$$
\mathfrak L_{i,j}(\boldsymbol{a},\boldsymbol{\lambda})=\mathfrak E_{i,j}\big (\pi(.;\boldsymbol{a});\mathcal V,\boldsymbol{u}\big )+\lambda_a \mathfrak B_{i,j}^{a}\big (\pi(.;\boldsymbol{a})\big )+\lambda_b \mathfrak B_{i,j}^{b}\big (\pi(.;\boldsymbol{a})\big ) 
$$
with $\boldsymbol{\lambda}=[\lambda_a,\lambda_b]^T$.

Due to the locality of the minimisation problem, we introduce the index $\ell=1,\cdots,|\mathcal V|$ and $C_\ell$, $M_\ell$ corresponds to cell $C_{i',j'}$ and centroid $M_{i',j'}$ with the local index respectively. Tensor $\boldsymbol{S}_{i,j}[\ell,i',j']$ of size $|\mathcal{V}|\times I\times J$ transforms the global indices $i',j'$ into the local index $\ell$ by setting $\boldsymbol{S}_{i,j}[\ell,i',j']=1$ if $\ell$ is the local index of cell $C_{i',j'}\in \mathcal V(C_{i,j})$, zero elsewhere. Consequently, vector $\boldsymbol{\varphi}_{i,j}=S_{i,j}\boldsymbol{\Phi}$ gathers the components $\boldsymbol{\varphi}_{i,j}[\ell]$ of the approximation with the local indexation.   

We introducing the matrix $\boldsymbol{A}_{i,j}$ of coefficients $\boldsymbol{A}_{i,j}[m,\ell]$, $m=1,\cdots,c_d$, $\ell=1,\cdots,|\mathcal V |$ with
$$
\boldsymbol{A}_{i,j}[m,\ell]=\frac{(M_\ell-M_{i,j})^{\gamma}}{\Delta x^{\gamma_x} \Delta y^{\gamma_y}},\quad  \gamma=\boldsymbol{\gamma}[m].
$$ 
Hence, the energy function reads
$$
\mathfrak E_{i,j}\big (\pi(.;\boldsymbol{a});\mathcal V,\boldsymbol{\varphi}_{i,j}\big )=
\sum_{\ell=1}^{|\mathcal V|} \frac{1}{2}\big (\boldsymbol{A}_{i,j}[.,\ell] \cdot \boldsymbol{a} -\boldsymbol{\varphi}_{i,j}[\ell]\big )^2,
\quad \boldsymbol{a}\in \mathbb R^{r_c},
$$
while the gradient reads
$$
\nabla_{\boldsymbol{a}} \mathfrak E\big (\pi(.;\boldsymbol{a});\mathcal V,\boldsymbol{\varphi}_{i,j}\big )=
\boldsymbol{A}_{i,j}^T \big (\boldsymbol{A}_{i,j}\boldsymbol{a} - \boldsymbol{\varphi}_{i,j} \big )=\boldsymbol{A}_{i,j}^T \boldsymbol{A}_{i,j}\boldsymbol{a}-\boldsymbol{A}_{i,j}^T\boldsymbol{S}_{i,j}\boldsymbol{\Phi}. 
$$

In the same way, the boundary condition reads for $r=a(i,j)$ and $r=b(i,j)$,
$$
\mathfrak B_{i,j}^r(\pi)=\alpha(P_r)\boldsymbol{a}\cdot \boldsymbol{\chi}_{i,j}(P_r)+\beta(P_r)\nabla_X (\boldsymbol{a}\cdot \boldsymbol{\chi}_{i,j})(P_r)\cdot n_{r}-g(P_r)=\boldsymbol{b}_{i,j}^r\cdot\boldsymbol{a}-g(P_r)
$$
with vector $\boldsymbol{b}_{i,j}^r\in\mathbb R^{c_d}$ given by the coefficients at $P_r=(x_r,y_r)$
$$
\boldsymbol{b}_{i,j}^r[m]=\frac{(x_r-x_i)^{\gamma_x}(y_r-y_j)^{\gamma_y}}{(\Delta x)^{\gamma_x}(\Delta y)^{\gamma_y}}
\Big (\alpha(P_r)+\beta(P_r)\frac{\gamma_x n_{r,x}}{(x_r-x_i)}+\beta(P_r)\frac{\gamma_y n_{r,y}}{(y_r-y_j)}\Big).
$$
$\gamma=\boldsymbol{\gamma}[m]$, $m=1,\cdots,c_d$.

Gathering the two vectors in the $c_d\times 2$ matrix $\boldsymbol{B}_{i,j}=\big [\boldsymbol{b}_{i,j}^a\ \boldsymbol{b}_{i,j}^b\big ]$, and let $\boldsymbol{g}_{i,j}=\big [g(P_a),\ g(P_b)\big ]^T$. The saddle point $(\boldsymbol{a}_{i,j},\boldsymbol{\lambda}_{i,j})$ satisfies the following $(c_d+2)\times(c_d+2)$ linear system
$$
\left [\begin{array}{cc} (\boldsymbol{A}_{i,j}^T \boldsymbol{A}_{i,j})  & \boldsymbol{B}_{i,j} \\ \boldsymbol{B}_{i,j}^T & 0 \end{array} \right ]
\left [\begin{array}{c} \boldsymbol{a}_{i,j} \\ \boldsymbol{\lambda}_{i,j} \end{array} \right ]=
\left [\begin{array}{c} \boldsymbol{A}_{i,j}^T \boldsymbol{S}_{i,j}\boldsymbol{\Phi} \\ \boldsymbol{g}_{i,j} \end{array} \right  ].
$$
Notice that all the small matrices $\boldsymbol{A}_{i,j}$, $\boldsymbol{B}_{i,j}$ are computed in a pre-processing stage together with matrix $(\boldsymbol{A}_{i,j}^T \boldsymbol{A}_{i,j})^{-1}$ that we use in the determination of the coefficients. 

\subsection{Validation of the ROD reconstruction}
Given a ghost cell $C_{i,j}$ of centroid $M_{i,j}$, given the boundary condition on the two collar points, we deduce the polynomial representative $\widehat \phi_{i,j}$ and then compute the value at point $M_{i,j}$. We perform all the reconstructions to set the values for the different layers of ghost-cells. Since the polynomial coefficients linearly depends on the values in the active cells, we deduce that the vector $\boldsymbol{\Phi}_{\GC}$ is an linear function of vector $\boldsymbol{\Phi}_{\IN}$ given by relation (\ref{eq:Roperator}).

To check the accuracy of Reconstruction Off-site Data procedure and assess the order method, we consider the function $\phi(x,y)=\exp(x+2y)$ and set the exact values in the active cells with $\phi_{ij}=\phi(x_i,y_j)$, $C_{i,j}\in \mathcal M_\Delta(\Omega)$. On the other hand, to fulfil the boundary condition at the collar points $P_k$, we manufacture the $g_k$ values such that $\mathfrak B(\phi)(P_k)=0$. The reconstruction procedure will then provide the values for the first, second and third layer of ghost cells we gather in vector $\boldsymbol{\Phi}_{\GC}$. 

\begin{rem}
To reduce the computational cost, we only evaluate the reconstructions for the ghost cells of the first layer and use the same polynomial to compute the extrapolations for the second and third layer.
\end{rem}

Errors are estimated with the $L^\infty$ norms for the first layer $\mathcal L_1=\mathcal L_1(\Omega_\Delta)$ with
$$
E^\infty_{\GC}=\max_{c_{i,j}\in \mathcal L_1}|\phi_{ij}-\phi(x_i,y_j)|.
$$
Table \ref{tab::ROD_val1} presents the errors and convergence order for the $\mathbb P_1$ reconstruction of the Dirichlet condition (first column), the $\mathbb P_1$ reconstruction of the Neumann reconstruction (second column) and the $\mathbb P_2$ reconstruction for the Neumann condition (third column). Notice that we lost one order of magnitude with the $\mathbb P_1$ reconstruction and Neumann condition but we manage to recover the optimal order using the $\mathbb P_2$ reconstruction. 

\begin{table}[ht]
{\notapolice
\begin{tabular}{|c|c|c|c|c|c|c|}
\hline
 & \multicolumn{2}{c|}{Dirichlet $\mathbb P_1$}& \multicolumn{2}{c|}{Neumann $\mathbb P_1$}    & \multicolumn{2}{c|}{Neumann $\mathbb P_2$}  \\ 
 \cline{2-7} 
\multirow{-2}{*}{I} & error    & order & error    & order & error    & order \\ \hline
80                  & 4.15e-03 &  ---  & 7.98e-02 &  ---  & 7.49e-04 &  ---  \\ \hline
160                 & 1.17e-03 & 1.83  & 4.42e-02 & 0.85  & 9.70e-05 & 2.95  \\ \hline
320                 & 3.26e-04 & 1.84  & 2.13e-02 & 1.05  & 1.22e-05 &  2.99  \\ \hline
\end{tabular}
\caption{\CapText ROD reconstruction convergence values for 3 combinations of boundary conditions and reconstruction polynomials: BC Dirichlet with $\mathbb P_1$, BC Neumann with $\mathbb P_1$ and BC Neumann with $\mathbb P_2$.}
\label{tab::ROD_val1} 
}
\end{table}

Similarly, we report in Table \ref{tab::ROD_val2} the error and convergence order for the $\mathbb P_3$ reconstruction both for Dirichlet and Neumann boundary condition, and the $\mathbb P_4$ reconstruction for the Neumann case. We observe that the Neumann condition does not suffer of the lack of accuracy as in the $\mathbb P_1$ case and already reach the optimal order.

\begin{table}[ht]
{\notapolice
\begin{tabular}{|c|c|c|c|c|c|c|}
\hline
     & \multicolumn{2}{c|}{Dirichlet $\mathbb P_3$} & \multicolumn{2}{c|}{Neumann $\mathbb P_3$}  & \multicolumn{2}{c|}{Neumann $\mathbb P_4$}  \\ 
\cline{2-7} 
\multirow{-2}{*}{I} & error    & order & error    & order & error    & order \\ \hline
80                  & 4.41e-05 & ---   & 4.32e-05 & ---   & 2.32e-06 & ---   \\ \hline
160                 & 3.51e-06 & 3.65  & 2.97e-06 & 3.86  & 1.39e-07 & 4.06  \\ \hline
320                 & 2.20e-07 & 4.00  & 1.85e-07 & 4.00  & 3.28e-09 & 5.41  \\ \hline
\end{tabular}
\caption{\CapText ROD reconstruction convergence values for 3 combinations of boundary conditions and reconstruction polynomials: BC Dirichlet with $\mathbb P_3$, BC Neumann with $\mathbb P_3$ and BC Neumann with $\mathbb P_4$.}
\label{tab::ROD_val2}
}
\end{table}

We present in Table \ref{tab::ROD-O6} the errors and convergence order for the $\mathbb P_5$ reconstruction both with the Dirichlet and Neumann condition. An additional benchmark is given for the $\mathbb P_6$ reconstruction with the Neumann condition. Indeed, for $I>160$ we reach the \texttt{double} precision capacity to handle the large condition number of the matrices involved in the polynomial coefficients' calculation. The right panel provides the errors for $I=60,80,100$ and show that we recover the the optimal order.

\begin{table}[ht]
{\notapolice
\begin{tabular}{|c|c|c|c|c|c|c|}
\hline
  & \multicolumn{2}{c|}{Dirichlet $\mathbb P_5$} & \multicolumn{2}{c|}{Neumann $\mathbb P_5$} & \multicolumn{2}{c|}{Neumann $\mathbb P_6$} \\ 
\cline{2-7} 
\multirow{-2}{*}{I} & error    & order& error    & order & error    & order \\ \hline
80                  & 6.57e-07 & ---  & 1.97e-07 &  ---  & 3.91e-08 &  ---  \\ \hline
160                 & 1.28e-08 & 5.68 & 3.72e-09 &  5.73 & 3.35e-09 & 3.54  \\ \hline
320                 & 5.13e-10 & 4.64 & 7.32e-11 & 5.67  & 2.61e-09 & 0.36  \\ \hline
\end{tabular}
\hskip 1em
\begin{tabular}{|c|l|l|}
\hline
\multirow{2}{*}{\textbf{I}} & \multicolumn{2}{c|}{Neumann $\mathbb P_6$} \\ 
\cline{2-3} 
                            & \multicolumn{1}{c|}{error} & \multicolumn{1}{c|}{order} \\ \hline
60                 & 1.89e-07                            & \multicolumn{1}{c|}{---}            \\ \hline
80                 & 3.91e-08                            & 5.48                                \\ \hline
100                & 9.94e-09                            & 6.14                                \\ \hline
\end{tabular}
\caption{\CapText ROD reconstruction convergence values for 3 combinations of boundary conditions and reconstruction polynomials: BC Dirichlet with $\mathbb P_5$, BC Neumann with $\mathbb P_5$ and BC Neumann with $\mathbb P_6$ (left panel),ROD reconstruction convergence values for BC Neumann with $\mathbb P_6$ using grids of smaller sizes (right panel).}
\label{tab::ROD-O6}
}
\end{table}


\section{The solver operator}

Given vector $\boldsymbol{\Phi}_{\GC}$, we seek vector $\boldsymbol{\Phi}_{\IN}$ deriving from a traditional finite difference scheme with Dirichlet condition at the ghost cells. It provides the linear solver operator (\ref{eq:Soperator}) by solving a linear system characterised by a very specific sparse matrix. Several algorithms are considered to take advantage of the structured mesh, namely parallel solvers that we shall evaluate in terms of efficiency and robustness.

\subsection{High-order finite difference schemes}
We adopt the standard finite difference schemes that we reproduce hereafter for the sake of consistency. We always consider centred schemes even for large cell P\'eclet number assuming that the solutions we are dealing with do not produce numerical instabilities. Of course, upwind scheme would be necessary in case of oscillations but this issue is out of the scope of the present paper.

The second-order scheme is achieved with the 3-points approximations
$$
\partial_x \phi(x_{i},y_{j})\approx \frac{-\phi_{i-1,j}+0\phi_{i,j}+\phi_{i+1,j}}{2\Delta x},\quad 
\partial_{xx} \phi(x_{i},y_{j})\approx \frac{\phi_{i-1,j}-2\phi_{i,j}+\phi_{i+1,j}}{(\Delta x)^2}.
$$
The fourth-order scheme derives from the 5-points approximations 
\begin{eqnarray*}
\partial_x \phi(x_{i},y_{j})   & \approx & \frac{ \phi_{i-2,j} -8\phi_{i-1,j}+0\phi_{i,j} + 8\phi_{i+1,j}-\phi_{i+2,j}}{12\Delta x}\\
\partial_{xx} \phi(x_{i},y_{j})& \approx & \frac{-\phi_{i-2,j}+16\phi_{i-1,j}-30\phi_{i,j}+16\phi_{i+1,j}-\phi_{i+2,j}}{12(\Delta x)^2}.
\end{eqnarray*}
We shall also consider the sixth-order scheme given by the 7-points approximations
\begin{eqnarray*}
\partial_x \phi(x_{i},y_{j})   & \approx & \frac{-\phi_{i-3,j} +9\phi_{i-2,j}-45\phi_{i-1,j}+0\phi_{i,j} + 45\phi_{i+1,j}- 9\phi_{i+2,j}+ \phi_{i+3,j}}
{60\Delta x}\\
\partial_{xx} \phi(x_{i},y_{j})& \approx & \frac{2\phi_{i-3,j}-27\phi_{i-2,j}+270\phi_{i-1,j}-490\phi_{i,j}+270\phi_{i+1,j}-27\phi_{i+2,j}+2\phi_{i+3,j}}
{180(\Delta x)^2}.
\end{eqnarray*}
We use the same discretisation for the $y$ direction. Notice that only the computational cells have to be evaluated since the values of the ghost cells were already evaluated in order to provide the necessary information to carry out the calculations. By solving the linear system, given the ghost cells vector $\boldsymbol{\Phi}_{\GC}$ and the right-hand side term, we obtain the linear operator $\boldsymbol{\Phi}_{\GC}\to \boldsymbol{\Phi}_{\IN}=\optS(\boldsymbol{\Phi}_{\GC};f)$. 

\subsection{Linear solvers and dimensional splitting}
Even enjoying a high degree of parallelisation, the Jacobi method converges too slowly when dealing with high conditioning number matrix and does not represent a satisfactory solution. The SOR technique converge faster but presents strong restrictions for a fully  parallelisation if one aims at using a large number of cores. For example, Red-Black ordering only uses two cores and the $m\times m$ block strategies creates strong overheads \cite{Evans1984}. 

Methods based on the residual computation such as the GMRES methods are strongly parallelisable by nature but memory access represents a limitation that strongly reduces its computational efficiency. 
Indeed, the Krylov based method involves the orthogonalisation procedure leading to the construction of full vectors and matrices together with a increasing computational cost. The biCGStab method is an interesting alternative since no additional storage is required but the computational cost is still significant (we require twice the evaluation of the residual) and the conditioning number is higher.


We here propose a more efficient method, fully paralellisable by construction, where data are consecutive in memory so what it takes advantage of the processor's cache hierarchy. We revisit the Alternate Direction Implicit method (ADI) proposed in the 60s by splitting the 2D system into a large number of 1D independent linear problem where the data is contiguous in memory, to leverage the cache access. Moreover, such a method does not require any additional storage or calculation (for instance the orthogonalisation). To this end, let consider the operators 
$$
L_x\phi =-\kappa\partial_{xx}\phi+u_x\partial_x\phi ,\qquad
L_y\phi =-\kappa\partial_{yy}\phi+u_y\partial_y\phi.
$$
We aim at seeking the solution $\overline{\phi}$ solution of the steady-state convection diffusion problem $0=L_x\overline{\phi}+L_y \overline{\phi}+f$ with Dirichlet boundary condition.  We slightly modify the equation we rewrite as a fix point problem by setting
$$
Id\,\overline{\phi}=Id\,\overline{\phi}+\tau L_x\overline{\phi}+\tau L_y \overline{\phi}+\tau f
$$
where $Id$ is the identity operator and $\tau>0$ a parameter.  At the discrete level, we denote by $\boldsymbol{L}_{\Delta x}$ and $\boldsymbol{L}_{\Delta y}$ the numerical discretisation matrices associated to operator $L_x$ and $L_y$ respectively while vector $\overline{\boldsymbol{\Phi}}$ is the solution of the linear problem
$$
\Id \, \overline{\boldsymbol{\Phi}}=\Id\,\overline{\boldsymbol{\Phi}}+\tau \boldsymbol{L}_{\Delta x} \overline{\boldsymbol{\Phi}}+\tau \boldsymbol{L}_{\Delta y} \overline{\boldsymbol{\Phi}}+\tau \boldsymbol{F}
$$
with $\boldsymbol{F}$ the discrete version of function $f$ and $\Id$ the identity matrix.
The ADI method is based on the construction of the following sequence $\big (\boldsymbol{\Phi}^{n}\big )_{n\in \mathbb N}$ of approximations given by
\begin{eqnarray*}
& &(\Id-\tau \boldsymbol{L}_{\Delta y})\boldsymbol{\Phi}^{n+1/2}=(\Id+\tau \boldsymbol{L}_{\Delta x})\boldsymbol{\Phi}^{n}+\tau \boldsymbol{F}\\
& &(\Id-\tau \boldsymbol{L}_{\Delta x})\boldsymbol{\Phi}^{n+1}=(\Id+\tau \boldsymbol{L}_{\Delta y})\boldsymbol{\Phi}^{n+1/2}
\end{eqnarray*}
where $\boldsymbol{\Phi}^{n+1/2}$ stands for an intermediate stage. 
At each stage $n$, we solve $I$ independent tri(penta, hepta)diagonal linear system ($x$-sweep) and $J$ independent tri(penta, hepta)diagonal linear system ($y$-sweep). The loop stops when the residual norm reach a prescribed tolerance. 

\begin{rem}
Higher order ADI would be considered to improve the iteration procedure. For example, the fourth-order ADI method proposed in \cite{KaZh2004} could be rewritten in the steady-state context.
\end{rem}
\subsubsection{Comparison with classical solvers}
We compare in tables \ref{tab::comparison_iterations_second_order} and \ref{tab::comparison_iterations_fourth_order} the ADI method with the GMRES and biCGStab for the second and fourth order finite difference method method. We consider the academic problem $-\Delta \phi=f$ with the exact solution $\phi(x,y)=\exp(x+2y)$ and the corresponding Dirichlet boundary condition. We solve the linear system until it reaches a residual lower than $10^{-12}$. 

We report the $L^2$ error and convergence order together with the number of iterations, \textit{i.e.} the number of calls to the residual computation using a $I\times I$ points grid. We obtain exactly the same errors and convergence order as expected. The number of iterations for the biCGStab is twice since we call two times the residual for each stages. The ADI method provides the lowest number of iterations and exhibit an excellent convergence. We do not present the computational times since they highly depend on the implementation of each method, the memory access, the cache memory, and the CPU Instructions Per Clock (IPC).  
\begin{table}[ht]
{\notapolice 
\begin{tabular}{|l|l|l|l|l|l|l|l|l|l|}
\hline
    & \multicolumn{3}{c|}{GMRES} & \multicolumn{3}{c|}{biCGStab} & \multicolumn{3}{c|}{ADI} \\ \hline
$I$ & err      & ord     & itr    & err      & ord     & itr    & err      & ord     & itr    \\ \hline \hline
10  & 1.01e-03 & ---     & 50     & 1.01e-03 & ---     & 104     & 1.01e-03 & ---    & 38     \\ \hline
20  & 2.82e-04 & 1.84    & 113    & 2.82e-04 & 1.84    & 232    & 2.82e-04 & 1.84    & 78     \\ \hline
40  & 7.47e-05 & 1.92    & 220    & 7.47e-05 & 1.92    & 460    & 7.47e-05 & 1.92    & 155    \\ \hline
80  & 1.93e-05 & 1.96    & 430    & 1.93e-05 & 1.96    & 876    & 1.93e-05 & 1.96    & 312    \\ \hline
120 & 8.65e-06 & 1.97    & 622    & 8.65e-06 & 1.97    & 1280    & 8.65e-06 & 1.97   & 467    \\ \hline
160 & 4.89e-06 & 1.98    & 788    & 4.89e-06 & 1.98    & 1696   & 4.89e-06 & 1.98    & 627    \\ \hline
\end{tabular}
}
\caption{\CapText Error, convergence order and number of iteration comparison between the solvers GMRES, biCGStab and ADI using a second order scheme for the Laplace equation.}
\label{tab::comparison_iterations_second_order}
\end{table}

\begin{table}[ht]
{\notapolice 
\begin{tabular}{|l|l|l|l|l|l|l|l|l|l|}
\hline
    & \multicolumn{3}{c|}{GMRES} & \multicolumn{3}{c|}{biCGStab} & \multicolumn{3}{c|}{ADI} \\ \hline
$I$ & err      & ord     & itr    & err      & ord     & itr    & err      & ord     & itr    \\ \hline \hline
10  & 9.32e-05 & ---     & 36     & 9.32e-05 & ---     & 80     & 9.32e-05 & ---     & 33     \\ \hline
20  & 9.46e-06 & 3.30    & 121    & 9.48e-06 & 3.30    & 232    & 9.48e-06 & 3.30    & 76     \\ \hline
40  & 7.36e-07 & 3.69    & 267    & 7.36e-07 & 3.69    & 480    & 7.36e-07 & 3.69    & 161    \\ \hline
80  & 5.11e-08 & 3.85    & 519    & 5.09e-08 & 3.85    & 964    & 5.09e-08 & 3.85    & 333    \\ \hline
120 & 1.02e-08 & 3.96    & 808    & 1.04e-08 & 3.92    & 1484   & 1.04e-08 & 3.92    & 503    \\ \hline
160 & 3.35e-09 & 3.92    & 1088   & 3.35e-09 & 3.94    & 1970   & 3.35e-09 & 3.94    & 665    \\ \hline
\end{tabular}
}
\caption{\CapText Error, convergence order and number of iteration comparison between the solvers GMRES, biCGStab and ADI using a fourth order scheme for the Laplace equation.}
\label{tab::comparison_iterations_fourth_order}
\end{table}
\newpage 

\subsubsection{Parallelism and computational efficiency}
The computational efficiency is deeply related to the built-in parallelism ability of the numerical method and its scalability. The main interest of the ADI solver is to split a 2D problem on a $I\times J$ grid into $I$ independent 1D problems on a $J$-points grid for the $x$-sweep (and the symmetric for the $y$-sweep). To this end, an OpenMP version of the code has been implemented in order to take advantage of the multi-threading capabilities of modern processors. 
Simulations have been carried out on a dual processor Intel Xeon E5-2650 v2 with 16 cores @2.6GHz and 64 GB of memory.

Table \ref{tab::speedup-O2} presents the time spent and respective speed-up from 1 to 16 cores and different $I\times I$ points grids for the centred second order scheme. We report a very good scaling when deploying up to 16 cores with a speed-up close to 9. We also note that the computational cost increases as $I^3$ while the grid size increases as $I^2$, hence the running time is proportional to the power $3/2$ of the number of unknowns.   
\begin{table}[ht]
{\notapolice
\begin{tabular}{|c|c|c|c|c|c|c|c|c|c|c|}
\hline
   & \multicolumn{5}{c|}{Time (s)}    & \multicolumn{5}{c|}{Speedup}         \\ 
\cline{2-11} 
\multirow{-2}{*}{I / \#cores} & 1 & 2 & 4 & 8 & 16 & 1 & 2 & 4 & 8 & 16 \\ \hline
256 &  1.42 & 0.74 & 0.44 & 0.27 &  0.2 & 1.00 & 1.92 & 3.23 & 5.26 & 7.10 \\ \hline
512 & 12.38 & 6.94 & 4.22 & 2.26 & 1.43 & 1.00 & 1.78 & 2.93 & 5.48 & 8.66 \\ \hline
1024&113.83 &64.41 &37.35 &20.89 &12.43 & 1.00 & 1.77 & 3.05 & 5.45 & 9.16 \\ \hline
\end{tabular}
}
\caption{\CapText ADI parallel implementation time and speedup results using up to 16 threads considering a 2nd order scheme.}
\label{tab::speedup-O2}
\end{table}

Tables \ref{tab::speedup-O4} and \ref{tab::speedup-O6} concern the fourth-order and the sixth-order scheme respectively. We remark that the speed-up is slightly better (close to 12 for 16 cores and the 6th-order scheme). We also notice that the running time is of the same order that $I^3$ for the 16 cores case, in line with the observation of the 2nd-order scheme. In particular for the 16 cores and $I=1024$ grid case, the computational time is 12.43 for the 2nd-order, 30.76 for the 4th-order ($\approx$two times), and only 38.76 ($\approx$three times) for the 6th-order. 

\begin{table}[ht]
{\notapolice
\begin{tabular}{|c|c|c|c|c|c|c|c|c|c|c|}
\hline
 & \multicolumn{5}{c|}{Time (s)}  & \multicolumn{5}{c|}{Speedup}  \\ 
 \cline{2-11} 
\multirow{-2}{*}{I / \#cores} & 1 & 2 & 4 & 8 & 16 & 1 & 2 & 4 & 8 & 16 \\ \hline
256 &  3.58 &  2.12 &  1.09 & 0.60 & 0.43 & 1.00 & 1.69 & 3.28 & 5.97 & 8.33 \\ \hline
512 & 38.81 & 19.74 & 12.02 & 6.34 & 3.73 & 1.00 & 1.97 & 3.23 & 6.12 &10.40 \\ \hline
1024&344.87 &193.34 &104.14 &55.83 &30.76 & 1.00 & 1.78 & 3.31 & 6.18 &11.21 \\ \hline
\end{tabular}
}
\caption{\CapText ADI parallel implementation time and speedup results using up to 16 threads considering a 4th order scheme.}
\label{tab::speedup-O4}
\end{table}

\begin{table}[ht]
{\notapolice
\begin{tabular}{|c|l|l|l|l|l|c|l|l|l|l|}
\hline
 & \multicolumn{5}{c|}{Time (s)} & \multicolumn{5}{c|}{Speedup} \\ 
\cline{2-11} 
\multirow{-2}{*}{I / \#cores} & 1 & 2 & 4 & 8 & 16 & 1 & 2 & 4 & 8 & 16 \\ \hline
256 &  5.77 &  2.65 &  1.58 & 0.86 & 0.58 & 1.00 & 2.18 & 3.65 & 6.71 &  9.95      \\ \hline
512 & 47.34 & 26.51 & 13.66 & 7.67 & 4.51 & 1.00 & 1.79 & 3.47 & 6.17 & 10.50     \\ \hline
1024&464.76 &247.99 &122.47 &68.71 &38.76 & 1.00 & 1.87 & 3.79 & 6.76 & 11.99     \\ \hline
\end{tabular}
}
\caption{\CapText ADI parallel implementation time and speedup results using up to 16 threads considering a 6th order scheme.}
\label{tab::speedup-O6}
\end{table}


\section{The fix-point solver}

The numerical solution of the convection diffusion problem (\ref{eq:ConvDiff})-(\ref{eq:BoundCond}) is provided by a sequence of general term $\boldsymbol{\Phi}^{n}_{\GC}$ given by the relation
$$
\boldsymbol{\Phi}^{n+1}_{\GC}=\optR(\optS(\boldsymbol{\Phi}^{n}_{\GC};f);g),\qquad \boldsymbol{\Phi}^{n+1}_{\IN}=\optS(\boldsymbol{\Phi}^{n+1}_{\GC};f).
$$
The sequence is built until we reach a satisfactory approximation of the fix-point solution $(\overline{\boldsymbol{\Phi}}_{\IN}, \overline{\boldsymbol{\Phi}}_{\GC})$. Since $\optR$ and $\optS$ are linear operators, the composition is also linear and reads
\begin{equation}\label{eq::fix_point_A_b}
\boldsymbol{\Phi}^{n+1}_{\GC}=A\boldsymbol{\Phi}^{n}_{\GC}-b   
\end{equation}
where $A$ is a full matrix of size the number of ghost cells. Moreover, let denote  $\Delta \boldsymbol{\Phi}^{n}_{\GC}= \boldsymbol{\Phi}^{n+1}_{\GC}-\boldsymbol{\Phi}^{n}_{\GC}$. Then, one has $\Delta \boldsymbol{\Phi}^{n+1}_{\GC}=A \Delta \boldsymbol{\Phi}^{n}_{\GC}$.

Since $A$ is not explicit, we use a matrix free procedure to provide an approximation of the fix point solution of the problem  $\overline{\boldsymbol{\Phi}}_{\GC}=A\overline{\boldsymbol{\Phi}}_{\GC}-b$.  It is of common knowledge that fix point method converges very slowly when matrix has some eigenvalues very close to one. We developed a new accelerator method to improve the solver's convergence rate.

\subsection{A preliminary analysis}
We consider in this section the very particular case where it exists an initial condition $\boldsymbol{\Phi}^{0}_{\GC}$ such that $\Delta \boldsymbol{\Phi}^{1}_{\GC}=\lambda \Delta \boldsymbol{\Phi}^{0}_{\GC}$ with $\lambda \in ]0,1[$, \textit{i.e.} vectors $\Delta \boldsymbol{\Phi}^{0}_{\GC}$ and $\Delta \boldsymbol{\Phi}^{0}_{\GC}$ are co-linear. Then the following proposition holds
\begin{prop}
For any $n\in \mathbb N$, one has 
$$
(a)\ \Delta \boldsymbol{\Phi}^{n}_{\GC}= \lambda^{n} \Delta \boldsymbol{\Phi}^{0}_{\GC}.
$$
Moreover, the exact solution is simply given by
$$
(b)\ \overline{\boldsymbol{\Phi}}_{\GC}=\boldsymbol{\Phi}^{0}_{\GC}+\frac{1}{1-\lambda}\Delta \boldsymbol{\Phi}^{0}_{\GC}.
$$
\end{prop}
\begin{pf}
Relation $(a)$ is obtained by applying $n$ times the operator. On the other hand, we write
\begin{equation}\label{eq::case_elementary_limit}
\boldsymbol{\Phi}^{n+1}_{\GC}-\boldsymbol{\Phi}^{0}_{\GC}=\sum_{\ell=0}^n \Delta \boldsymbol{\Phi}^{\ell}_{\GC}=\sum_{\ell=0}^n \lambda^{\ell} \Delta \boldsymbol{\Phi}^{0}_{\GC}=\frac{1-\lambda^{n+1}}{1-\lambda}\Delta \boldsymbol{\Phi}^{0}_{\GC}
\end{equation}
We then deduce that sequence $\boldsymbol{\Phi}^{n+1}_{\GC}$ converges and relation (\ref{eq::fix_point_A_b}) gives
$$
\lim_{n\to \infty} \boldsymbol{\Phi}^{n+1}_{\GC}=A\lim_{n\to \infty} \boldsymbol{\Phi}^{n}_{\GC}-b.
$$
We conclude that the limit is the fix point $\overline{\boldsymbol{\Phi}}_{\GC}$. Passing to the limit in relation (\ref{eq::case_elementary_limit}), provides relation $(b)$.\qed 
\end{pf}
The preliminary study indicates that the co-linearity property between two successive iterations is a key to compute a better approximation of the solution, avoiding all the intermediate stages and saving a lot of computational effort. Of course, in general case, we do not have such an ideal situation but when co-linearity is almost achieved, we shall take advantage of the property as presented in the next section. 
 
\subsection{Fix point accelerators}

We drop the subscript $\GC$ for the sake of simplicity and define the two following quantities:  
$$
\lambda^{(n)}=\frac{|\Delta \boldsymbol{\Phi}^{n}|}{|\Delta \boldsymbol{\Phi}^{n-1}|}, \qquad 
C^{(n)}=\cos(\theta^{(n)})=\frac{\Delta \boldsymbol{\Phi}^{n}\cdot \Delta \boldsymbol{\Phi}^{n-1}}{|\Delta \boldsymbol{\Phi}^{n}|  |\Delta \boldsymbol{\Phi}^{n-1}| },
$$
where $|u|$ and $u\cdot v$ stand for the Euclidean norm and inner product between vectors $u$ and $v$.
From the definitions, we have the following result.
\begin{prop}
Let $\boldsymbol{\Phi}^{0}$ be the initial condition.  Then the following decomposition holds
$$
\Delta \boldsymbol{\Phi}^{1}=C^{(1)}\lambda^{(1)}\Delta \boldsymbol{\Phi}^{0}+\Delta \boldsymbol{\Psi}^{0}
$$
with $\Delta \boldsymbol{\Phi}^{0}\cdot \Delta \boldsymbol{\Psi}^{0}=0$. Moreover, one has
$$
\Delta \boldsymbol{\Phi}^{n}=\big (C^{(1)}\lambda^{(1)}\big )^n \Delta \boldsymbol{\Phi}^{0}+
\sum_{\ell=0}^{n-1} \big (C^{(1)}\lambda^{(1)}\big )^\ell A^{n-1-\ell}\Delta \boldsymbol{\Psi}^{0}.
$$
Finally, the fix point solution $\overline{\boldsymbol{\Phi}}$ satisfies the relation
\begin{equation}\label{eq::decomposition_fixpoint}
\overline{\boldsymbol{\Phi}}=\boldsymbol{\Phi}^{0}+\frac{1}{1-C^{(1)}\lambda^{(1)}} \Delta \boldsymbol{\Phi}^{0}+
\frac{(Id-A)^{-1}}{1-C^{(1)}\lambda^{(1)}} \Delta \boldsymbol{\Psi}^{0}.
\end{equation}
\end{prop}
\begin{pf}
From the definitions of $C^{(1)}$ and $\lambda^{(1)}$, we have
$$
\frac{\Delta \boldsymbol{\Phi}^{1}\cdot \Delta \boldsymbol{\Phi}^{0} }{|\Delta \boldsymbol{\Phi}^{0}|  |\Delta \boldsymbol{\Phi}^{0}| }=
\frac{\Delta \boldsymbol{\Phi}^{1}\cdot \Delta \boldsymbol{\Phi}^{0} }{|\Delta \boldsymbol{\Phi}^{1}|  |\Delta \boldsymbol{\Phi}^{0}| }\times \frac{|\Delta \boldsymbol{\Phi}^{1}|}{|\Delta \boldsymbol{\Phi}^{0}|}=C^{(1)}\lambda^{(1)}=\tau.
$$
Hence defining vector $\boldsymbol{\Psi}^{0}=\Delta \boldsymbol{\Phi}^{1}-\tau \Delta \boldsymbol{\Phi}^{0}$, the orthogonality property holds by construction.

We prove the second relation by induction. We compute
\begin{eqnarray*}
\Delta \boldsymbol{\Phi}^{n+1}&=&A\Delta \boldsymbol{\Phi}^{n}=A \left [ \tau^n \Delta \boldsymbol{\Phi}^{0}+\sum_{\ell=0}^{n-1} \tau^\ell A^{n-1-\ell}\Delta \boldsymbol{\Psi}^{0} \right ] \\
&=&\tau^n  A\Delta \boldsymbol{\Phi}^{0}+\sum_{\ell=0}^{n-1} \tau^\ell A^{n-1-\ell+1}\Delta \boldsymbol{\Psi}^{0}\\
&=&\tau^{n+1} \Delta \boldsymbol{\Phi}^{0}+\tau^n \Delta \boldsymbol{\Psi}^{0}+\sum_{\ell=0}^{n-1} \tau^\ell A^{n-\ell}\Delta \boldsymbol{\Psi}^{0} \\
&=&\tau^{n+1} \Delta \boldsymbol{\Phi}^{0}+
\sum_{\ell=0}^{n} \tau^\ell A^{n-\ell}\Delta \boldsymbol{\Psi}^{0}.
\end{eqnarray*}
To prove the last relation, we write
$$
\Delta \boldsymbol{\Phi}^{0}=A\boldsymbol{\Phi}^{0}-b-\boldsymbol{\Phi}^{0}=A\boldsymbol{\Phi}^{0}-\boldsymbol{\Phi}^{0}+\overline{\boldsymbol{\Phi}}-A\overline{\boldsymbol{\Phi}}=(\Id-A)(\overline{\boldsymbol{\Phi}}-\boldsymbol{\Phi}^{0}).
$$
Multiplying the relation with $(1-\tau)$ gives
\begin{eqnarray*}
(1-\tau)(\Id-A)(\overline{\boldsymbol{\Phi}}-\boldsymbol{\Phi}^{0})&=&(1-\tau)\Delta \boldsymbol{\Phi}^{0}\\
&=&\Delta \boldsymbol{\Phi}^{0}-\Delta \boldsymbol{\Phi}^{1}+\Delta \boldsymbol{\Phi}^{1}-\tau \Delta \boldsymbol{\Phi}^{0}\\
&=& \Delta \boldsymbol{\Phi}^{0}-\Delta \boldsymbol{\Phi}^{1}+\Delta \boldsymbol{\Psi}^{0}\\
&=& (\Id -A) \Delta \boldsymbol{\Phi}^{0}+\Delta \boldsymbol{\Psi}^{0}
\end{eqnarray*}
Assuming that matrix $(\Id-A)$ in non-singular, we obtain
$$
(1-\tau)(\overline{\boldsymbol{\Phi}}-\boldsymbol{\Phi}^{0})=\Delta \boldsymbol{\Phi}^{0}+(\Id-A)^{-1}\Delta \boldsymbol{\Psi}^{0}.
$$
Hence by dividing with the quantity $1-\tau$, we deduce the relation (\ref{eq::decomposition_fixpoint}).\qed 
\end{pf}
Relation (\ref{eq::decomposition_fixpoint}) states that the fix point solution is decomposed into a principal with $\boldsymbol{\Phi}^{0}$ and a complementary orthogonal part. Note that the accelerator parameter $\frac{1}{1-C^{(1)}\lambda^{(1)}}$ turns to be very large when $C^{(1)}\lambda^{(1)}$ are close to one (co-linearity). We propose several improvements of the fix point method based on that remark. 

\subsubsection{First acceleration algorithm}
Assume that at stage $n$ we know the approximation $\boldsymbol{\Phi}^{n}$. 
\begin{enumerate}
\item  We compute two successive steps noting 
$$
\boldsymbol{\Phi}^{n,\star}=A\boldsymbol{\Phi}^{n}-b,\quad \boldsymbol{\Phi}^{n,\star\star} =A\boldsymbol{\Phi}^{n,\star}-b,
$$ 
and the increments
$$
\Delta \boldsymbol{\Phi}^{n} =\boldsymbol{\Phi}^{n,\star}-\boldsymbol{\Phi}^{n}, \quad \Delta \boldsymbol{\Phi}^{n,\star} =\boldsymbol{\Phi}^{n,\star\star}-\boldsymbol{\Phi}^{n,\star}.
$$
\item We compute the indicators 
$$
\lambda^{(n,\star)}=\frac{|\Delta \boldsymbol{\Phi}^{n,\star}|}{|\Delta \boldsymbol{\Phi}^{n}|}, \qquad 
C^{(n,\star)}=\cos(\theta^{(n,\star)})=\frac{\Delta \boldsymbol{\Phi}^{n,\star}\cdot \Delta \boldsymbol{\Phi}^{n} }{|\Delta \boldsymbol{\Phi}^{n,\star}|  |\Delta \boldsymbol{\Phi}^{n}| }.
$$
\item The new approximation is then given by truncation of the second term in relation (\ref{eq::decomposition_fixpoint}). 
\begin{equation}\label{eq::first_algo}
\boldsymbol{\Phi}^{n+1}=\boldsymbol{\Phi}^{n} + \frac{1}{1-C^{(n,\star)}\lambda^{(n,\star)}} \Delta \boldsymbol{\Phi}^{n}
\end{equation}
\end{enumerate}

\subsubsection{Second acceleration algorithm}
To improve the performance, we design a second algorithm by using the orthogonality property  
$$
\Delta \boldsymbol{\Phi}^{0,\star}= \tau \Delta \boldsymbol{\Phi}^{0}+\Delta \boldsymbol{\Psi}^{0}
$$
where we set $\tau=C^{(0,\star)}\lambda^{(0,\star)}$ for the sake of notation. We then have the following proposition
\begin{prop}
Let $\boldsymbol{\Phi}^{0}$ the initial condition. The following decomposition holds
\begin{equation}\label{eq::second_decomposition_fixpoint}
\overline{\boldsymbol{\Phi}}= \boldsymbol{\Phi}^{0}+\frac{1}{1-\tau} \ \frac{\Delta \boldsymbol{\Phi}^{0,\star} }{\tau}+
\frac{(Id-A)^{-1}}{1-\tau} A \Delta \boldsymbol{\Psi}^{0} + \frac{\Delta \boldsymbol{\Psi}^{0}}{\tau}.
\end{equation}
\end{prop}
\begin{pf}
Inserting relation 
$$\Delta \boldsymbol{\Phi}^{0}=\frac{\Delta \boldsymbol{\Phi}^{0,\star}-\Delta \boldsymbol{\Psi}^{0}}{\tau}$$ 
into relation (\ref{eq::decomposition_fixpoint}) provides
 \begin{eqnarray*}
\overline{\boldsymbol{\Phi}}&=&\boldsymbol{\Phi}^{0}+\frac{1}{1-\tau} \, \frac{\Delta \boldsymbol{\Phi}^{0,\star}-\Delta \boldsymbol{\Psi}^{0}}{\tau}+\frac{(Id-A)^{-1}}{1-\tau} \Delta \boldsymbol{\Psi}^{0}\\
&=&\boldsymbol{\Phi}^{0}+\frac{1}{1-\tau} \, \frac{\Delta \boldsymbol{\Phi}^{0,\star} }{\tau}+
\frac{1}{1-\tau} \Big ( (Id-A)^{-1}\Delta \boldsymbol{\Psi}^{0} -\frac{\Delta \boldsymbol{\Psi}^{0}}{\tau}\Big )\\
&=& \boldsymbol{\Phi}^{0}+\frac{1}{1-\tau} \, \frac{\Delta \boldsymbol{\Phi}^{0,\star} }{\tau}+
\Big ( (Id-A)^{-1} -\frac{Id}{\tau}\Big )\frac{\Delta \boldsymbol{\Psi}^{0}}{1-\tau} \\
&=&\boldsymbol{\Phi}^{0}+\frac{1}{1-\tau} \, \frac{\Delta \boldsymbol{\Phi}^{0,\star} }{\tau}+
\Big ( (Id-A)^{-1} -Id +\frac{1-\tau}{\tau}Id\Big )\frac{\Delta \boldsymbol{\Psi}^{0}}{1-\tau}  \\
&=&\boldsymbol{\Phi}^{0}+\frac{1}{1-\tau} \, \frac{\Delta \boldsymbol{\Phi}^{0,\star} }{\tau}+
\Big ( A(Id-A)^{-1} +\frac{1-\tau}{\tau}Id\Big ) \frac{\Delta \boldsymbol{\Psi}^{0}}{1-\tau}  \\
&=&\boldsymbol{\Phi}^{0}+\frac{1}{1-\tau} \, \frac{\Delta \boldsymbol{\Phi}^{0,\star} }{\tau}+
\frac{(Id-A)^{-1}}{1-\tau} A \Delta \boldsymbol{\Psi}^{0} +
\frac{\Delta \boldsymbol{\Psi}^{0}}{\tau}.\\
\end{eqnarray*}
\end{pf}

Since $\tau \in]0,1[$, we have $\frac{1}{(1-\tau)\tau}>\frac{1}{1-\tau}$. Hence the second approximation  
$$
\boldsymbol{\Phi}^{1}=\boldsymbol{\Phi}^{0} + \frac{1}{1-C^{(0,\star)}\lambda^{(0,\star)}} \times \frac{\Delta \boldsymbol{\Phi}^{0,\star}}{C^{(0,\star)}\lambda^{(0,\star)}}
$$
obtained by eliminating the orthogonal contribution provides a better estimation than the one given by equation (\ref{eq::first_algo}). 

We extend the formulae to any stage $n$ assuming that we know the approximation $\boldsymbol{\Phi}^{n}$. The new algorithm reads:
\begin{enumerate}
\item  We compute two successive steps noting 
$$
\boldsymbol{\Phi}^{n,\star}=A\boldsymbol{\Phi}^{n}-b,\quad \boldsymbol{\Phi}^{n,\star\star} =A\boldsymbol{\Phi}^{n,\star}-b,
$$ 
and the increments
$$
\Delta \boldsymbol{\Phi}^{n} =\boldsymbol{\Phi}^{n,\star}-\boldsymbol{\Phi}^{n}, \quad \Delta \boldsymbol{\Phi}^{n,\star} =\boldsymbol{\Phi}^{n,\star\star}-\boldsymbol{\Phi}^{n,\star}.
$$
\item We compute the indicators 
$$
\lambda^{(n,\star)}=\frac{|\Delta \boldsymbol{\Phi}^{n,\star}|}{|\Delta \boldsymbol{\Phi}^{n}|}, \qquad 
C^{(n,\star)}=\cos(\theta^{(n,\star)})=\frac{\Delta \boldsymbol{\Phi}^{n,\star}\cdot \Delta \boldsymbol{\Phi}^{n} }{|\Delta \boldsymbol{\Phi}^{n,\star}|  |\Delta \boldsymbol{\Phi}^{n}| }.
$$
\item The new approximation is then given by truncation of the second term in relation (\ref{eq::second_decomposition_fixpoint}). 
\begin{equation}\label{eq::second_algo}
\boldsymbol{\Phi}^{n+1}=\boldsymbol{\Phi}^{n} + \frac{1}{1-C^{(n,\star)}\lambda^{(n,\star)}} \times \frac{\Delta \boldsymbol{\Phi}^{n,\star}}{C^{(n,\star)}\lambda^{(n,\star)}}.
\end{equation}
\end{enumerate}

\subsection{Accelerator performance}
To assess the performance of the second accelerator, we solve the convection diffusion problem $U\cdot\nabla\phi-\kappa \Delta \phi=f$ taking $\phi(x,y)=\exp(x+2y)$ and manufacturing the adequate right-hand side source term $f$. Domain $\Omega$ is an open disk of radius $r=0.8$ while $\Lambda$ is the square $[-1,1]^2$. We perform the convergence iterative process until we satisfy the stopping criterion $\vert \boldsymbol{\Phi}^{n}-\boldsymbol{\Phi}^{n+1}\vert <\varepsilon_T$ with $\varepsilon_T$ the tolerance. All the tests are carried out with a  Intel Core i7-6700HQ @2.60GHz and 16GB of memory.

We recall that the global solver involves two nested loops: the outer one for the fix point problem that modifies the boundary condition via the ghost cells and the inner one where we solve the ADI problem for given ghost cell values. We present in Table \ref{tab::FixPoint_acc_DirichleT} the number of outer loop iterations (named ROD iterations) and the cumulative inner loop iterations (named ADI iterations) for the case of Dirichlet condition. The left panel provides the numbers for a tolerance of $\varepsilon_T=10^{-13}$ while the right panel gives the same information for a larger tolerance $\varepsilon_T=10^{-11}$. We identify the acceleration activation by the label {\notapolice acc} and we use {\notapolice no acc} correspond to the method with no acceleration. 

We obtain an important reduction of the computational effort due to a dramatic reduction of the number of iterations. The case $I=320$ shows that we cut by a thrid the computational time. We note that the tolerance does not significantly impact the quantification (running time and number of iterations). 
\begin{table}[ht]
{\notapolice
\begin{tabular}{|c|c|c|c|c|}
\hline
\multirow{2}{*}{I} & \multicolumn{2}{c|}{ROD iters  (ADI iters)} & \multicolumn{2}{c|}{Time (s)} \\ 
\cline{2-5} 
    & no acc & acc & no acc &  acc \\ \hline
80  & 54 (4379)  & 33 (1214) & 0.76  & 0.34 \\ \hline
160 & 68 (7727)  & 34 (1964) & 4.02  & 1.24 \\ \hline
320 & 53 (15132) & 45 (4605) & 28.02 & 9.01 \\ \hline
\end{tabular}\hskip 0.5em
\begin{tabular}{|c|c|c|c|c|}
\hline
\multirow{2}{*}{I} & \multicolumn{2}{c|}{ROD iters  (ADI iters)} & \multicolumn{2}{c|}{Time (s)} \\ 
\cline{2-5} 
   & no acc & acc  & no acc & acc \\ \hline
 80 & 39 (4320) & 29 (1134) & 0.73 & 0.30 \\ \hline
160 & 39 (7295) & 30 (1858) & 3.74 & 1.21 \\ \hline
320 & 44 (15073)& 34 (4019) &27.69 & 8.17 \\ \hline
\end{tabular}
}
\caption{\CapText Convergence rate (number of iterations and time spent) with and without the accelerator: Dirichlet boundary conditions  ($\varepsilon_T=10^{-13}$ left panel), ($\varepsilon_T=10^{-11}$ right panel).}
\label{tab::FixPoint_acc_DirichleT}
\end{table}

\begin{figure}[ht]
\includegraphics[width=0.95\textwidth]{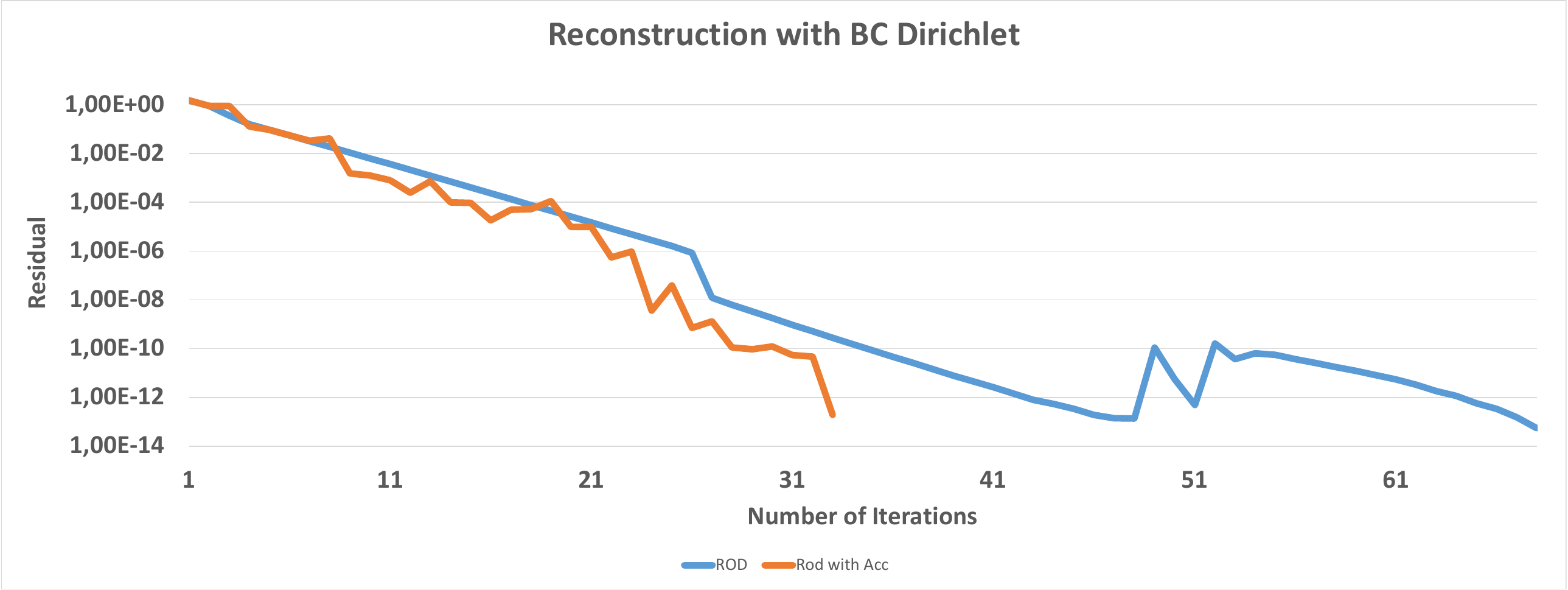}
\centering
\caption{\CapText Histogram of the residual of the fix point method with Dirichlet boundary conditions with and without accelerator. ($I=160$)}
\label{fig:acc_convergence_dirichlet}
\end{figure}

We plot in Figure \ref{fig:acc_convergence_dirichlet} the histogram of the residual between two successive solutions both with and without acceleration. We remark that the two curves are roughly similar up to an error of $\varepsilon_T=10^{-8}$ and then the acceleration provides better convergence while the traditional fix-point method presents large oscillations around $\varepsilon_T=10^{-11}$.

We perform a similar benchmark but with the Robin condition case taking $\alpha=\beta=1$. We report in Table  \ref{tab::FixPoint_acc_RobiN} the iterations' number and times that strongly differ from the Dirichlet case. Indeed, the normal derivative contribution strongly controls the convergence rate. With $80\times 80$ cells, the computational effort is divided by $4$ for a tolerance of $\varepsilon_T=10^{-13}$ and $6$ for $\varepsilon_T=10^{-11}$ but the noticeable effect of the accelerator is fully underlined with $320\times 320$ where the computational effort is cut by $24$.  and a number of cumulative ADI iterations divided by 36. Another noticeable point is that the execution time increases with a factor $10$ when $I$ doubles while the factor is only four with the acceleration procedure. For example, the finer mesh with no acceleration takes $40$ times the duration of the simulation with the accelerator.  
\begin{table}[ht]
{\notapolice
\hskip -2em 
\begin{tabular}{|c|c|c|c|c|}
\hline
\multirow{2}{*}{I} & \multicolumn{2}{c|}{ROD iters  (ADI iters)} & \multicolumn{2}{c|}{Time (s)} \\ 
\cline{2-5} 
 & no acc & acc & no acc & acc \\ \hline
 80 & 4.9k (524k)  & 2.5k (33k)  & 107.53   & 25.82  \\ \hline
160 & 8.4k (1.9M) & 4.3k (77k)  & 1.0k  & 112.05 \\ \hline
320 & 19.3k (6.6M)& 6.1k (180k) & 12.5k & 532.04 \\ \hline
\end{tabular}
\hskip 0.5em 
\begin{tabular}{|c|c|c|c|c|}
\hline
\multirow{2}{*}{\textbf{I}} & \multicolumn{2}{c|}{ROD iters  (ADI iters)} & \multicolumn{2}{c|}{Time (s)} \\ 
\cline{2-5} 
 & no acc    &  acc & no acc &  acc   \\ \hline
80   & 3.8 (514k)  & 1.5k (20k)  & 96.16    & 15.64  \\ \hline
160  & 7.4k (1.9M) & 2.5k (44k)  & 975.98   & 25.77  \\ \hline
320  & 14k (6.4M)& 3.6k (109k) & 12k & 325.96 \\ \hline
\end{tabular}
}
\caption{\CapText Convergence rate (number of iterations and time spent) with and without the accelerator: Robin boundary conditions  ($\varepsilon_T=10^{-13}$ left panel), ($\varepsilon_T=10^{-11}$ right panel). Notice that $10k$ means $10000$ and $10M$ represents ten millions.}
\label{tab::FixPoint_acc_RobiN}
\end{table}

To reinforce our comments, we display the residual norm versus the number of iterations both with and without the acceleration. Unlike the Dirichlet case, the slopes are quite different and the acceleration procedure clearly brings important gains. We note the effect of the predictor that produces some high frequency oscillations but with a global decrease of the residual. 
\begin{figure}[ht]
\includegraphics[width=0.95\textwidth]{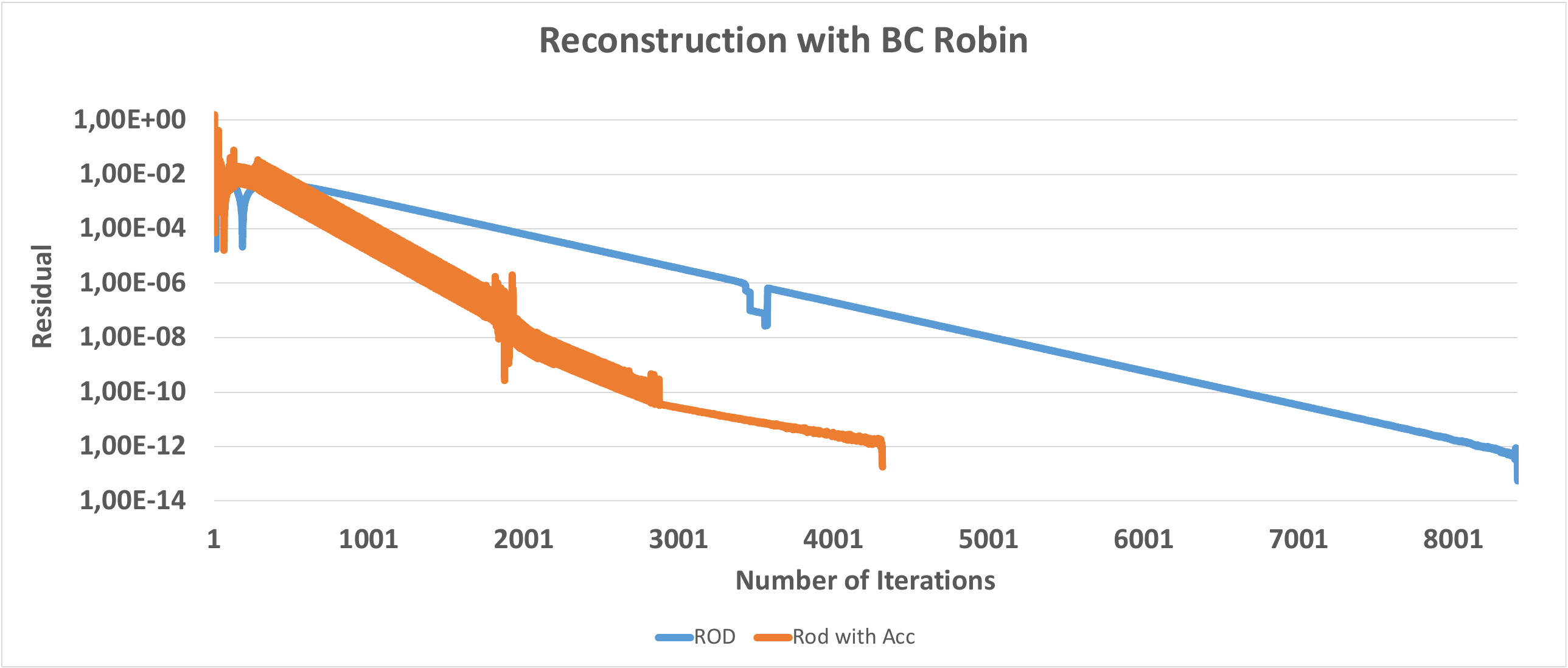}
\centering
\caption{\CapText Histogram of the residual of the fix point method with Robin boundary conditions with and without accelerator. ($I=160$)}
\label{fig:rod_160_robin}
\end{figure}


\section{Numerical tests}

\subsection{Convection diffusion order check}
To assess the convergence order, we propose two benchmarks with manufactured solutions and curved shape domains. We also consider different kinds of boundary conditions and show that we always recover the optimal convergence order.
\subsubsection{Annulus domain} 
\begin{figure}[ht]
\includegraphics[width=0.5\textwidth]{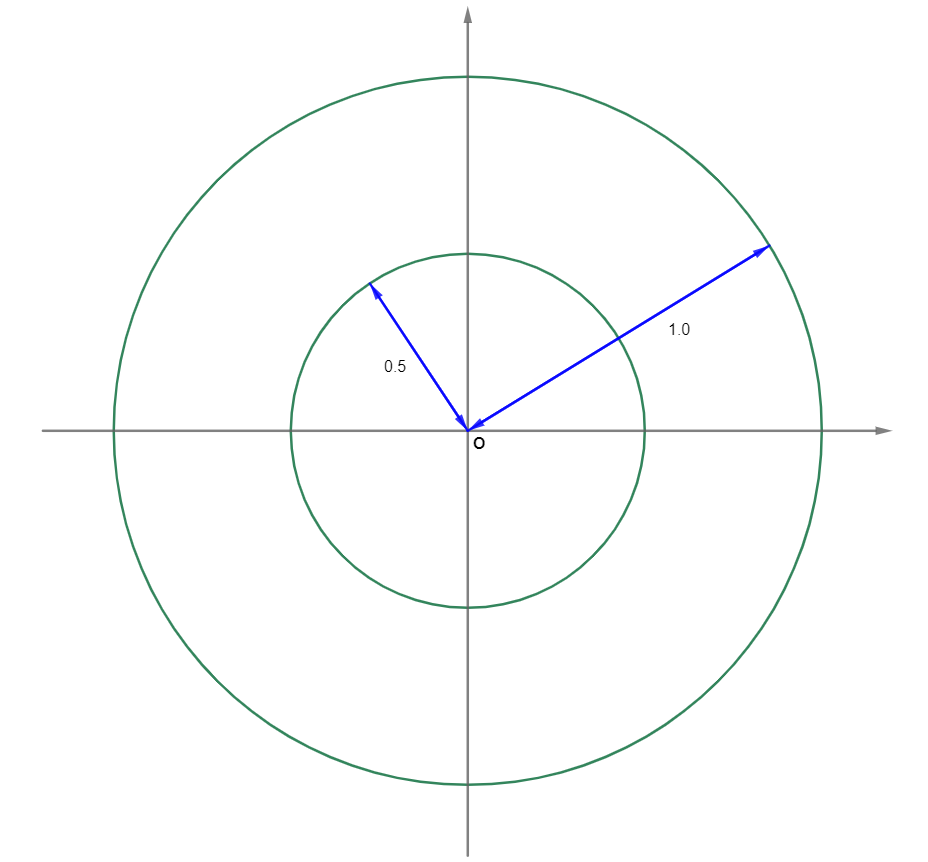}
\centering
\caption{Annulus shape.}
\label{fig:annulus_shape}
\end{figure}

The physical domain is an annulus of inner radius $R_i=0.5$ and outer radius $R_o=1.0$ as displays in Fig. \ref{fig:annulus_shape}. The extended domain $\Lambda$ is the square $[-1.25,1.25]\times[-1.25,1.25]$ to catch both the active and ghost cells for all the reconstruction orders. Function $\phi(x,y)=\frac{a}{2}\ln(x^2+y^2)+b$ is solution of the Laplace equation $\Delta \phi=0$ inside the physical domain. The boundary conditions of Dirichlet or Neumann condition are constant values on $R_i$ and $R_e$ by construction.

To assess the convergence order, we compute the numerical approximation with several meshes and compare it to the exact solution. We consider two situations whether we use Dirichlet-Dirichlet condition or Neumann-Dirichlet condition (inner and outer border respectively). We report in Table \ref{tab::Tests-annulus-o2}  the $L^\infty$-error and convergence order both to the D-D and the D-N cases using the $\mathbb P_1$ reconstruction for the Dirichlet side and  the $\mathbb P_2$ polynomial for the Neumann side. We recover the optimal second-order convergence rate in the two cases.

\begin{table}[ht]
{\notapolice
\begin{tabular}{|c|c|c|c|c|}
\hline
\multirow{2}{*}{I} 
& \multicolumn{2}{c|}{\begin{tabular}[c]{@{}c@{}}Dirichlet - Dirichlet \\ (O2 $\mathbb P_1$)\end{tabular}} 
& \multicolumn{2}{c|}{\begin{tabular}[c]{@{}c@{}}Neumann -Dirichlet \\ (O2 $\mathbb P_2$)\end{tabular}} \\ 
\cline{2-5} 
    & error & order & error  & order \\ \hline
120 & 5.30e-04 & ---  & 7.79e-04 & ---  \\ \hline
240 & 1.47e-04 & 1.85 & 2.25e-04 & 1.79 \\ \hline
320 & 7.64e-05 & 2.27 & 1.34e-04 & 1.80 \\ \hline
\end{tabular}
}
\caption{\CapText Errors and convergence rate for the 2nd order approximation (right panel) on an annulus shape with constant boundary conditions: Dirichlet on the inner and outer circle; Neumann in the inner circle and Dirichlet for the outer circle.}
\label{tab::Tests-annulus-o2}
\end{table} 

Tables \ref{tab::Tests-annulus-o4o6} provides the errors and convergence rates for the fourth-order (left panel) and sixth-order (right panel) reconstructions with the specific correction for the Neumann boundary condition. We report that the optimal order is achieved once again. 

\begin{table}[ht]
{\notapolice
\hskip -2em
\begin{tabular}{|c|c|c|c|c|}
\hline
\multirow{2}{*}{I} 
& \multicolumn{2}{c|}{\begin{tabular}[c]{@{}c@{}}Dirichlet - Dirichlet \\ (O4 $\mathbb P_3$)\end{tabular}} 
& \multicolumn{2}{c|}{\begin{tabular}[c]{@{}c@{}}Neumann -Dirichlet \\ (O4 $\mathbb P_4$)\end{tabular}} \\ 
\cline{2-5} 
    &  error  & order&  error  & order \\ \hline
120 & 2.26e-05& ---  & 1.86e-04& ---   \\ \hline
240 & 1.41e-06& 4.00 & 1.61e-05& 3.53  \\ \hline
320 & 5.05e-07& 3.57 & 5.53e-06& 3.71  \\ \hline
\end{tabular}
\hskip 0.5em
\begin{tabular}{|c|c|c|c|c|}
\hline
\multirow{2}{*}{I} 
& \multicolumn{2}{c|}{\begin{tabular}[c]{@{}c@{}}Dirichlet - Dirichlet \\ (O6 $\mathbb P_5$)\end{tabular}} 
& \multicolumn{2}{c|}{\begin{tabular}[c]{@{}c@{}}Neumann -Dirichlet \\ (O6 $\mathbb P_6$)\end{tabular}} \\ 
\cline{2-5} 
    & error & order & error & order  \\ \hline
120 & 5.70e-06&  --- & 6.10e-05& ---  \\ \hline
240 & 1.30e-07& 5.45 & 2.05e-06& 4.90 \\ \hline
320 & 2.95e-08& 5.16 & 4.53e-07& 5.25 \\ \hline
\end{tabular}
}
\caption{\CapText Errors and convergence rate for the 4th order (left panel) and 6th order approximation (right panel) on an annulus shape with constant boundary conditions: Dirichlet on the inner and outer circle; Neumann in the inner circle and Dirichlet for the outer circle.}
\label{tab::Tests-annulus-o4o6}
\end{table} 

 \subsubsection{Non-polynomial domain}
\begin{figure}[ht]
\includegraphics[width=0.9\textwidth]{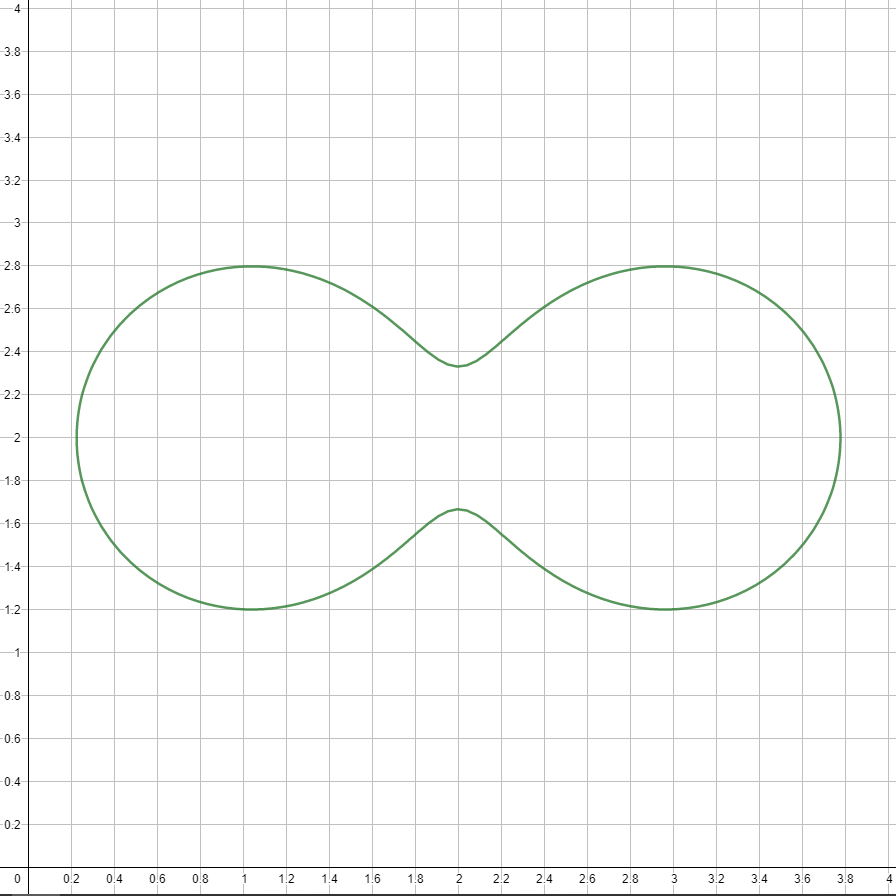}
\centering
\caption{\CapText The non polygonal domain and the background grid.}
\label{fig:biscuit_shape}
\end{figure}

 We proceed with a non-polynomial domain where the boundary does not derive from the zero-level of a polynomial function. The domain $\Omega$ is depicted in figure \ref{fig:biscuit_shape}  inside the larger domain $\Lambda=[0,4]\times [0,4]$. The manufactured function is once again $\phi(x,y)=\exp(x+2y)$.

Numerical simulations are carried out for the Dirichlet or the Robin boundary condition. We report in Table  \ref{tab::Tests-biscuit-o2} the second-order and fourth-order method with the $\mathbb P_1$ and $\mathbb P_3$ polynomial reconstruction obtained by the ROD method with the additional degree when dealing with the Robin boundary condition. Notice that the reported convergence order is slightly lower than the optimal order since we use the $L^\infty$. Fortunately, we recover the full order with the $L^1$ norm not provided here for the 2nd and 4th order.

\begin{table}[ht]
{\notapolice
\begin{tabular}{|c|c|c|c|c|}
\hline
\multirow{2}{*}{I} 
& \multicolumn{2}{c|}{\begin{tabular}[c]{@{}c@{}}Dirichlet \\ (O2 $\mathbb P_1$)\end{tabular}} 
& \multicolumn{2}{c|}{\begin{tabular}[c]{@{}c@{}}Robin\\ (O2 $\mathbb P_2$)\end{tabular}} \\ 
\cline{2-5} 
 & error & order & error & order \\ \hline
120 & 3.63e-03& ---  & 1.05e-01& ---  \\ \hline
240 & 1.04e-03& 1.80 & 2.79e-02& 1.91 \\ \hline
320 & 6.19e-04& 1.80 & 1.57e-02& 2.00 \\ \hline
\end{tabular}
\hskip 0.5em
\begin{tabular}{|c|c|c|c|c|}
\hline
\multirow{2}{*}{I} 
& \multicolumn{2}{c|}{\begin{tabular}[c]{@{}c@{}}Dirichlet \\ (O4 $\mathbb P_3$)\end{tabular}} 
& \multicolumn{2}{c|}{\begin{tabular}[c]{@{}c@{}}Robin\\ (O4 $\mathbb P_4$)\end{tabular}} \\ 
\cline{2-5} 
    & error & order  & error & order \\ \hline
120 & 2.46e-05& ---  & 4.43e-04& --- \\ \hline
240 & 1.63e-06& 3.92 & 2.19e-05& 4.34\\ \hline
320 & 5.31e-07& 3.90 & 7.40e-06& 3.77\\ \hline
\end{tabular}
}
\caption{\CapText Error and convergence rate for the 2nd- (left panel) and 4th- order (right panel) solution with Dirichlet and Robin boundary condition in $L^\infty$-norm.}
\label{tab::Tests-biscuit-o2}
\end{table} 

\begin{table}[ht]
{\notapolice
\begin{tabular}{|c|c|c|c|c|}
\hline
\multirow{2}{*}{I} 
& \multicolumn{2}{c|}{\begin{tabular}[c]{@{}c@{}}Dirichlet \\ (O6 $\mathbb P_5$)\end{tabular}} 
& \multicolumn{2}{c|}{\begin{tabular}[c]{@{}c@{}}Robin\\ (O6 $\mathbb P_6$)\end{tabular}} \\ 
\cline{2-5} 
  & error & order & error & order \\ \hline
120 & 2.48e-07& --- & 9.06e-06& --- \\ \hline
240 & 7.10e-09& 5.13 & 1.91e-07& 5.57 \\ \hline
320 & 1.36e-09& 5.74 & 1.21e-07& 1.59 \\ \hline
\end{tabular}
\hskip 0.5em
\begin{tabular}{|c|c|c|}
\hline
\multirow{2}{*}{I} 
& \multicolumn{2}{c|}{\begin{tabular}[c]{@{}c@{}}Robin \\ (O6 P6)\end{tabular}} \\ \cline{2-3} 
  & error & order    \\ \hline
120 & 2.13e-06& ---  \\ \hline
240 & 5.59e-08& 5.25 \\ \hline
320 & 8.73e-09& 6.45 \\ \hline
\end{tabular}
}
\caption{\CapText Error and convergence rate for the 6th-order solution with Dirichlet and Robin boundary condition: $L^\infty$-norm (left) and $L^1$-norm (right).}
\label{tab::Tests-biscuit-o6}
\end{table} 
The sixth-order method is assessed and errors are reported in Table \ref{tab::Tests-biscuit-o6} for the $L^\infty$ (left panel) and the $L^1$-norm (right panel). We reach to the machine precision capacity with the largest mesh and convergence order is no longer available. We then evaluate the order with coarser meshes and the $L^1$-norm to highlight that we obtain the optimal order.

\subsection{Non-rotational flow in a nozzle with obstacles}
\begin{figure}[ht]
\includegraphics[width=0.9\textwidth]{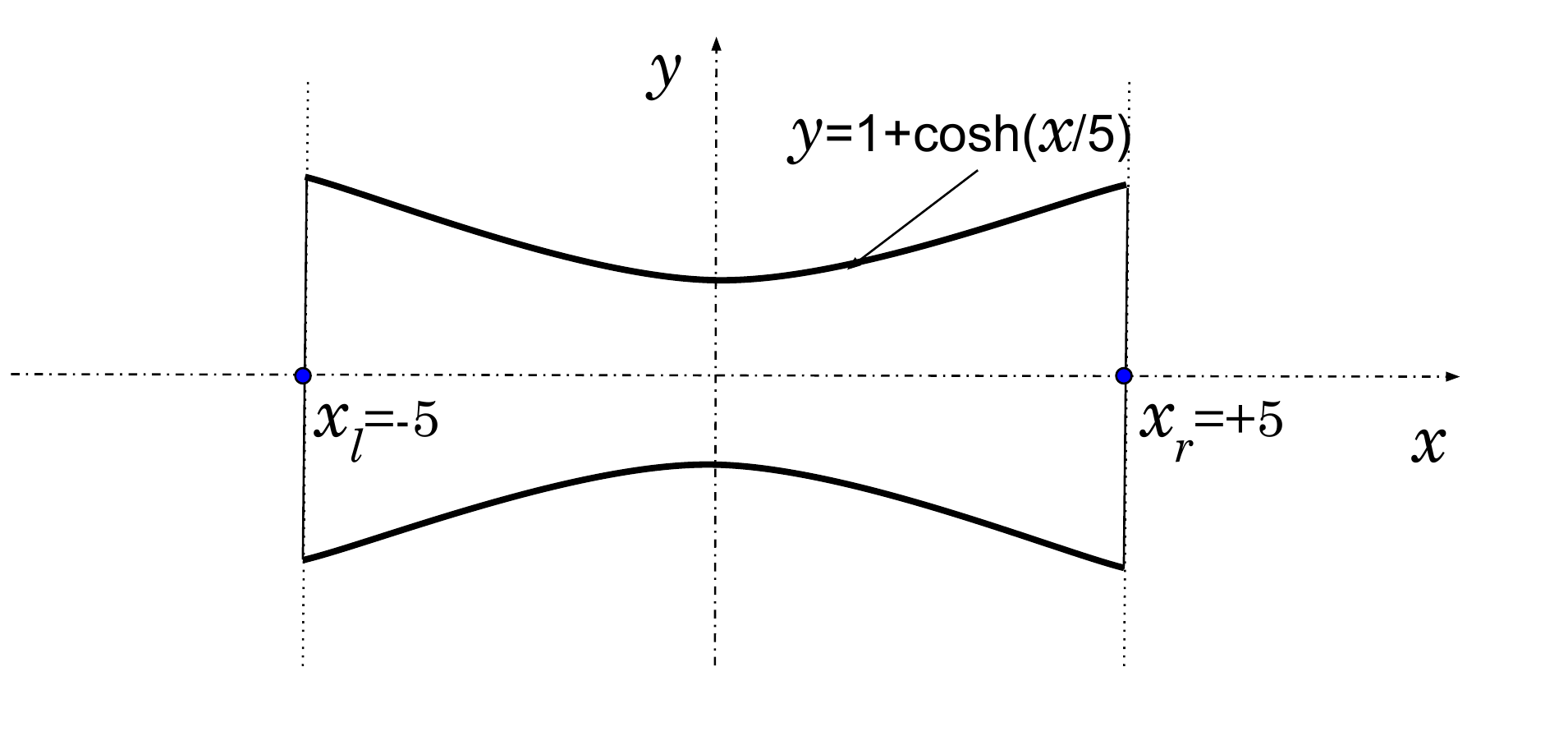}
\centering
\caption{Nozzle-shape domain.}
\label{fig::nozzle_domain}
\end{figure}

We considered a 2D symmetric nozzle-shape domain where the upper and lower sides are given by a $y=1+\cosh(x/5)$ and $y=-1-\cosh(x/5)$ while the left and right side are situated at $x_l=-5$ and $x_r=5$ respectively (see figure (\ref{fig::nozzle_domain}). we have carried out the simulation with several successive nested meshes $\mathcal M^r$, $r=1,2,3,4$ corresponding to $I=40$, $120$, $360$, $1080$ respectively and $J=I$ such that the centroids of the coarsest mesh $r=1$ are also centroids of the finer meshes $r=2,3,4$.

\subsubsection{Sanity-check benchmark}
We consider the Laplace operator $-\Delta \phi=f$ with $\phi=\ln(1.0+x^{2}+3y^{2}))$ and $f$ the corresponding source term.
We prescribe Dirichlet condition on the left and right side and Neumann condition up and down with the help of the exact solution.

We report in Table \ref{tab::conv_ln} the $L^\infty$-error and convergence order for the 2nd-, 4th- and 6th-order method. The highest order is rather degraded for the finer meshes. The large stencil required by the 6th-order together with the four corners of the domain lead to high conditioning number matrices in the ROD reconstruction that limits its convergence. Indeed, some geometrical configurations give rise to a very poorly conditioning linear system which strongly impact the highest order convergence.  

\begin{table}[ht]
{\notapolice
\begin{tabular}{|c|c|c|c|c|c|c|}
\hline
 & \multicolumn{2}{c|}{2nd Order}  & \multicolumn{2}{c|}{4th Order}  & \multicolumn{2}{c|}{6th Order}  \\ \hline
I & err & ord &  err  &  ord  &  err  &  ord  \\ \hline
40  & 1.27e-01 & ---  & 2.05e-02 & ---  & 1.15e-01 & --- \\ \hline
120 & 1.52e-02 & 1.93 & 6.18e-04 & 3.19 & 1.42e-04 & 6.10 \\ \hline
360 & 1.55e-03 & 2.08 & 6.60e-06 & 4.13 & 2.41e-06 & 3.71 \\ \hline
1080& 1.68e-04 & 2.02 & 1.54e-07 & 3.42 & 2.39e-05 & ---  \\ \hline
\end{tabular}
}
\caption{\CapText Errors and converge orders for the 2nd-, 4th- and 6th-order schemes.}
\label{tab::conv_ln}
\end{table} 

Convergence in meshes is also evaluated and Table \ref{tab::conv_mesh_ln} shows the successive approximations $\phi^r(M)$ at the node $M=(2.1125; 0.8125)$ for the different nested meshes $r=1,2,3,4$. The difference is evaluated by the successive differences $\delta^r=|\phi^r(M)-\phi^{r-1}(M)|$, $r=2,3,4$ while the convergence rate is given by $\alpha^r=\ln(\delta^{r-1}/\delta^r)/\ln(3)$, $r=3,4$.
The difference between two successive solutions is an ersatz of the error assessment without accessing the exact solution while the ratio evaluates the convergence in meshes of the numerical method. We observe the expected convergence rate for $I=360$, $1080$ cases but with a strong default for the 6th-order in line with the convergence rate given in  Table  \ref{tab::conv_ln}

\begin{table}[ht]
{\notapolice
\begin{tabular}{|c|c|c|c|c|c|c|c|c|c|}
\hline
 & \multicolumn{3}{c|}{2nd Order} & \multicolumn{3}{c|}{4th Order}  & \multicolumn{3}{c|}{6th Order} \\ \hline
I & value & $\delta$ & $\alpha$ & value & $\delta$ & $\alpha$ & value & $\delta$ & $\alpha$ \\ \hline
40  & 1.956119 & ---      & ---  & 2.019278 & ---      & ---  & 1.926048 & ---      & ---  \\ \hline
120 & 2.000286 & 4.42E-02 & ---  & 2.007161 & 1.21E-02 & ---  & 2.007355 & 8.13E-02 & ---  \\ \hline
360 & 2.006614 & 6.33E-03 & 1.77 & 2.007289 & 1.28E-04 & 4.14 & 2.007292 & 6.28E-05 & 6.52 \\ \hline
1080& 2.007219 & 6.05E-04 & 2.14 & 2.007291 & 1.67E-06 & 3.95 & 2.007275 & 1.68E-05 & 1.20 \\ \hline
\end{tabular}
}
\caption{\CapText Convergence rate in mesh for the $\phi=\ln(1.0+x^{2}+3y^{2}))$ with the 2nd-, 4th- and 6th-order schemes at point $P_{1}$.}
\label{tab::conv_mesh_ln}
\end{table}

\subsubsection{Irrotational flow in a nozzle-shape domain}
\begin{figure}[ht]
\includegraphics[width=0.5\textwidth]{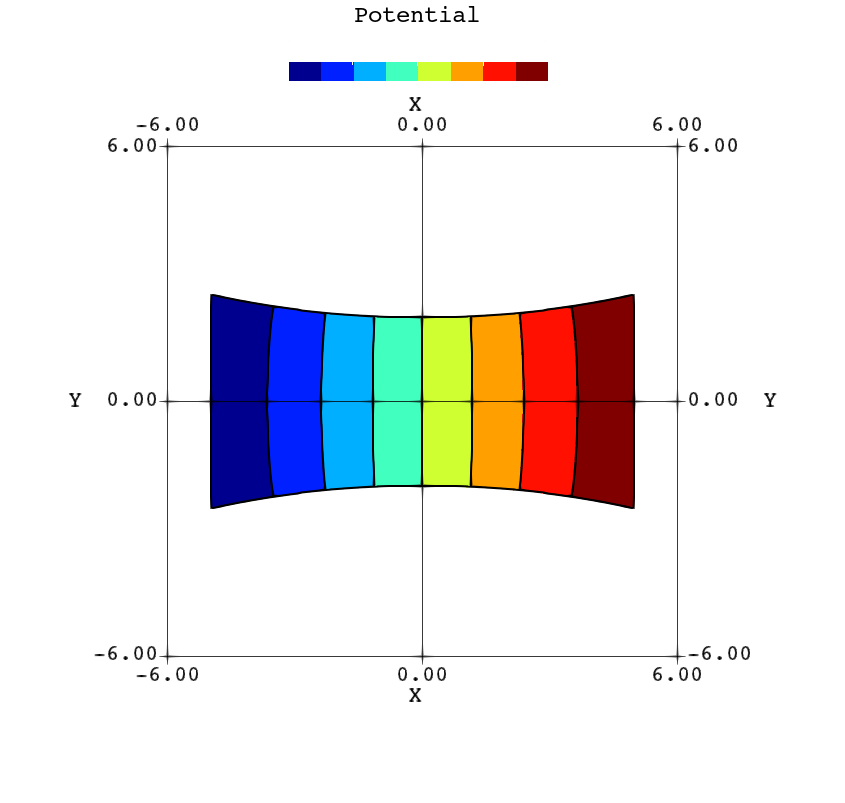}\hskip 1em
\includegraphics[width=0.45\textwidth]{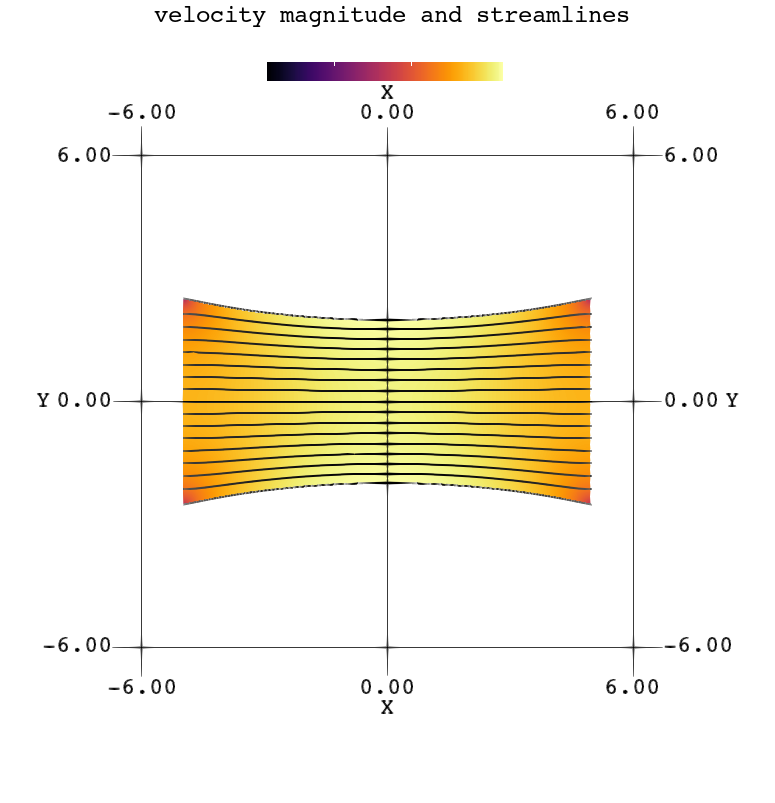}
\centering
\caption{Potential function in the Nozzle with 8 uniform isovalues from -1 to 1 (left). Streamlines and velocity magnitude (right).}
\label{fig:nozzle_1}
\end{figure}

An irrotational flow is described by a potential function $\phi$ such that $U=\nabla \phi$ together with the free divergence condition $\nabla . U=0$. We prescribe the Dirichlet boundary conditions $\phi(-10,y)=-1$ and $\phi(10,y)=1$ on the left and right side while the top and bottom surfaces satisfy the wall condition $\nabla \phi \cdot n=0$. We perform the computation with the four meshes and report in Table \ref{tab::conv_mesh_irrotational} the value, differential and rate for the 2nd-, 4th- and 6th order methods. Figure \ref{fig:nozzle_1} displays the potential function and the velocity field.

We note that the rates are not the expected one, in particular the 6th-order method reduce to the second-order of convergence. Nevertheless, several aspects should be mentioned. The absolute error is quite small with respect to the former case while the two solutions roughly range in same interval of value. We suggest that the solution symmetries are responsible for a very low error even with a coarse mesh that masks the expected convergence (see the next case for complementary arguments). 
\begin{table}[ht]
{\notapolice
\begin{tabular}{|c|c|c|c|c|c|c|c|c|c|}
\hline
 & \multicolumn{3}{c|}{2nd Order} & \multicolumn{3}{c|}{4th Order}  & \multicolumn{3}{c|}{6th Order} \\ \hline
I & value & $\delta$ & $\alpha$ & value & $\delta$ & $\alpha$ & value & $\delta$ & $\alpha$ \\ \hline
40  & 0.445472 & ---      & ---  & 0.446880 & ---      & ---  & 0.447604 & ---      & ---  \\ \hline
120 & 0.445766 & 2.93E-04 & ---  & 0.446032 & 8.48E-04 & ---  & 0.446009 & 1.59E-03 & ---  \\ \hline
360 & 0.445948 & 1.82E-04 & 0.43 & 0.445987 & 4.50E-05 & 2.67 & 0.445985 & 2.38E-05 & 3.83 \\ \hline
1080& 0.445978 & 3.02E-05 & 1.64 & 0.445983 & 3.74E-06 & 2.26 & 0.445983 & 2.42E-06 & 2.08 \\ \hline
\end{tabular}
}
\caption{\CapText Convergence rate in mesh for the irrotational flow: 2nd-, 4th- and 6th-order schemes for a point $\mathbb P_{1}$.}
\label{tab::conv_mesh_irrotational}
\end{table} 

\subsubsection{Irrotational flow with obstacles}
\begin{figure}[ht]
\centering
\includegraphics[width=0.45\textwidth]{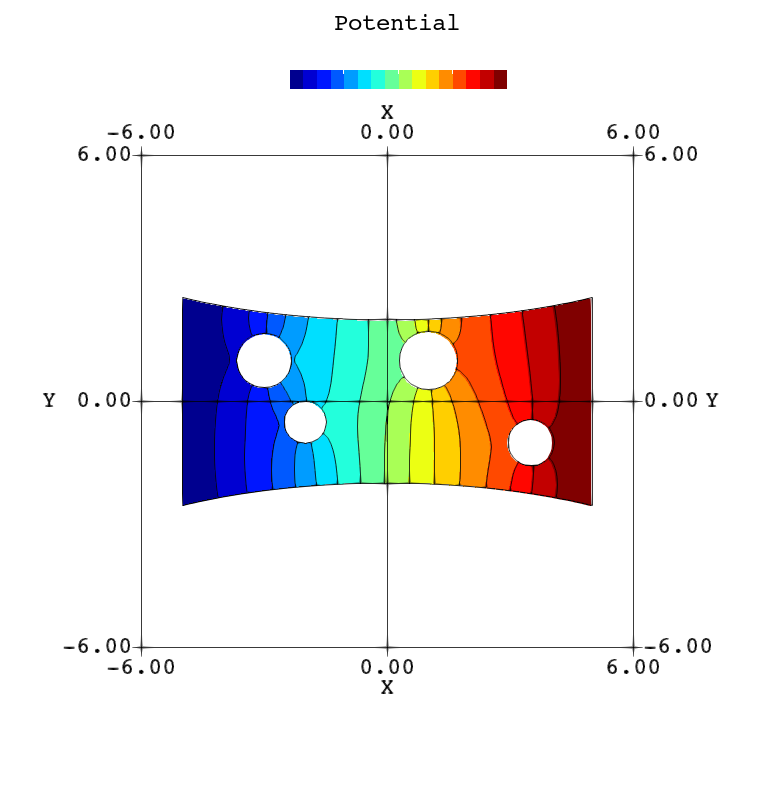}\hskip 1em
\includegraphics[width=0.45\textwidth]{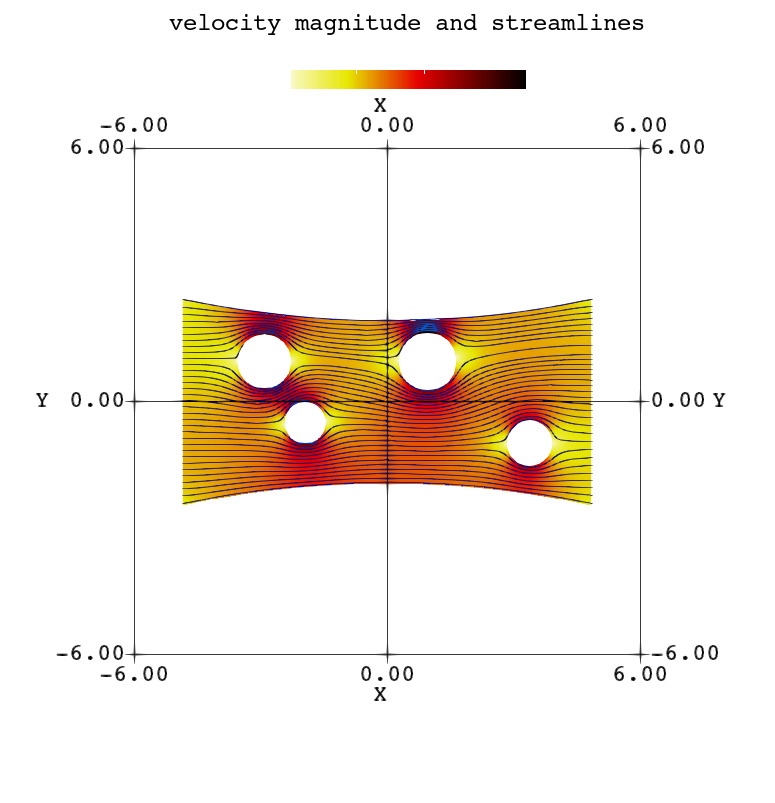}
\caption{Potential function in the Nozzle with obstacles and 16 uniform isovalues from -1 to 1 (left). Velocity magnitude and streamlines (right).}
\label{fig:nozzle_2}
\end{figure}

We domain is filled with three obstacles where we prescribe the solid wall boundary. Figure \ref{fig:nozzle_2} (left) depicts the flow and and right panel check the tangential property of the velocity on the boundary. 

We assess the convergence in grid at point $M=(1.1375; 0.8125)$ and we present both the differences and ratios for the successive meshes. We note that the absolute differences $\delta$ on the previous table are lower of two magnitudes that the ones obtained with the same domain without obstacle.  Moreover, we recover the correct order for the 2nd- and 4th-order. We then collect more evidences that the previous case enjoys some strong reduction of the error even with the coarse meshes that mask the order. One more time, the 6th-order of convergence suffer of bad conditioning stencils and almost reach the fourth-order of convergence.

\begin{table}[ht]\label{tab::conv_mesh_obstacle}
{\notapolice
\begin{tabular}{|c|c|c|c|c|c|c|c|c|c|}
\hline
 & \multicolumn{3}{c|}{2nd Order} & \multicolumn{3}{c|}{4th Order}  & \multicolumn{3}{c|}{6th Order} \\ \hline
I & value & $\delta$ & $\alpha$ & value & $\delta$ & $\alpha$ & value & $\delta$ & $\alpha$ \\ \hline
40  & 0.133168 & ---      & ---  & 0.455705 & ---      & ---  & 0.485895 & ---      & ---  \\ \hline
120 & 0.195705 & 6.25E-02 & ---  & 0.232193 & 2.24E-01 & ---  & 0.208364 & 2.78E-01 & ---  \\ \hline
360 & 0.219412 & 2.37E-02 & 0.88 & 0.219284 & 1.29E-02 & 2.60 & 0.219345 & 1.10e-02 & 2.94 \\ \hline
1080& 0.219199 & 2.13E-04 & 4.29 & 0.219197 & 8.77E-05 & 4.54 & 0.219176 & 1.69e-04 & 3.80 \\ \hline
\end{tabular}
}
\caption{\CapText Convergence rate in mesh for the irrotational fluid with obstacles: 2nd-, 4th- and 6th-order schemes for a point $\mathbb P_{1}$.}
\end{table} 

\section{Conclusions}
Very high-order schemes on Cartesian grid with arbitrary regular geometries is a critical issue to achieve high quality approximations while taking advantage of the computational efficiency of the data structures. We propose a general method to handle Robin conditions (including Dirichlet and Neumann as a particular case) relying on local information transfer (boundary location and condition) into a polynomial representation we use to fill the ghost cells. The simplicity of the boundary representation (no analytical representation is necessary, no orthogonal projection is performed) enables a high versatility of the technique to handle complex boundaries 

Additionally, the method is presented as the coupling of two independent black-boxes, by splitting up the boundary problem with the interior problem. Consequently, on the one hand, we take advantage of the Cartesian grid by developing a ADI-like dimensional splitting that transform a full 2D problem into a multitude of independent 1D problems we solve in parallel. On the other hand, each polynomial reconstruction is determined independently and boundary conditions are prescribed in an universal manner. A global efficiency of the method in the many-core context is achieved where scalability and speed-up are almost optimal.

\section*{Acknowledgements}
The authors acknowledge the financial support by FEDER -- Fundo Europeu de Desenvolvimento Regional, through COMPETE 2020 -- Programa Operacional Fatores de Competitividade, and the National Funds through FCT -- Fundação para a Ciência e a Tecnologia, project No. POCI-01-0145-FEDER-028118, PTDC/MAT-APL/28118/2017.\\
This work was partially  financially supported by: Project POCI-01-0145-FEDER-028247 - funded by FEDER funds through COMPETE2020 - Programa Operacional Competitividade e Internacionalização (POCI) and by national funds (PIDDAC) through FCT/MCTES.\\
This work was supported by the Portuguese Foundation for Science and Technology (FCT) in the framework of the Strategic Funding UIDB/04650/2020.
\bibliographystyle{elsarticle-num} 
\bibliography{main.bib}
\end{document}